\documentclass[]{interact}

\usepackage{epstopdf}
\usepackage[caption=false]{subfig}

\usepackage[numbers,sort&compress]{natbib}
\bibpunct[, ]{[}{]}{,}{n}{,}{,}
\renewcommand\bibfont{\fontsize{10}{12}\selectfont}
\makeatletter
\def\NAT@def@citea{\def\@citea{\NAT@separator}}
\makeatother

\usepackage{xcolor}

\theoremstyle{plain}

\theoremstyle{definition}

\theoremstyle{remark}

\begin{document}


\title{The Supernova 1987A system and its recent evolution - a review}

\author{
\name{Michael~J. Barlow\textsuperscript{a}\thanks{CONTACT M.~J. Barlow. Email: m.barlow@ucl.ac.uk}}
\affil{\textsuperscript{a} Dept. of Physics \& Astronomy, University College London, Gower Street, \\ London WC1E 6BT, UK}
}

\maketitle

\begin{abstract}
Supernova~1987A was the closest supernova event to be observed in nearly 400 years. 
The outflowing ejecta from the explosion continues to interact with extended 
circumstellar material and with the equatorial ring (ER) in the triple ring system, 
while observations of the system have continued across the whole electromagnetic 
spectrum. This review mainly focuses on works published over the past ten years on 
Supernova~1987A and its remnant. These include (a) submillimetre, infrared and X-ray 
studies of molecules, ions and dust in the ejecta, and (b) infrared, optical and 
X-ray studies of dust and ionized gas in the ER and in the surrounding circumstellar 
medium, including their time evolution as the ER is shocked and eroded by the impact 
of high velocity ejecta. Since 2022, the {\em James Webb Space Telescope} has become 
available for high angular resolution infrared observations of Supernova 1987A and 
has made significant contributions to both (a) and (b) above. Its discovery of 
redshifted narrow-line emission from multiple ion species located at the centre of 
the ejecta strongly requires the presence there of either a cooling hot neutron star, 
or a pulsar wind nebula, to power the emission.

\textcolor{blue}{\bf To appear in `Contemporary Physics'.}


\end{abstract}


\section{Introduction}
Supernova 1987A (SN~1987A) in the Large Magellanic Cloud (LMC) exploded at 7:35 UT February 23rd 1987. Its progenitor star was quickly identified as the 12th magnitude early B-type blue supergiant Sk~-69~202. Sanduleak's objective-prism spectrum, acquired in the 1960's \citep{Sanduleak1970}, was the only one ever taken of this star. The presence of hydrogen lines in the spectrum of SN~1987A classified it as a Type~II or core-collapse supernova (CCSN). As the nearest and brightest optical supernova since Kepler's supernova of 1604\footnote{The rate at which CCSNe should occur in the LMC has been estimated to be once every 200~years \citep{Tammann1994}, who also estimated the rate for our own Milky Way (MW) galaxy to be once every 47$\pm$12 years, based on comparisons with similar spiral galaxies. A CCSN rate for the MW of one every 53$^{+70}_{-20}$ years has been estimated \citep{Diehl2006} from $\gamma$-ray observations of radioactive $^{26}$Al produced by massive stars. More recently, the MW CCSN rate has been estimated to be one every 220$^{+130}_{-40}$ years, from a census of massive star formation rates in the MW \citep{Quintana2025}. Since suitable neutrino detectors have been in operation since 1982, we know that no CCSN events can have occurred in our Galaxy since then.}, it immediately attracted intense interest and continues to do so, effectively conferring on it the status of an archetypal CCSN, despite some notable peculiarities.

The proximity of SN~1987A, combined with its recent occurrence, has enabled it to be studied from the outset at high signal to noise ratios across the entire electromagnetic spectrum using modern technologies, both ground-based and spaceborne. It provided the first detection of neutrinos from outside the solar system, initiating the era of neutrino astronomy. CCSNe are the principal sources of many of the elements found in the Universe today, including the products of nuclear fusion, from oxygen to the iron group, along with the elements beyond the iron peak created by the rapid (r-process) addition of free neutrons to nuclei during the core-collapse explosion event. In the case of SN~1987A, we can uniquely measure not only the abundances of elements from different layers of the exploded star, but also abundance ratios for many isotopes, using millimetre-wave, infrared, X-ray and $\gamma$-ray telescopes that can probe deep into the post-explosion ejecta. 

The advent of large scale extragalactic supernova surveys and their follow-ups has led to
the realisation that many CCSNe show evidence of interactions with circumstellar material, with onset timescales ranging from a few tens of days to several years. SN~1987A has its own circumstellar ring structures (Section 2) while, unlike other similarly young SNe, its proximity enables spatially resolved studies of the shock interactions of the outflowing ejecta with the circumstellar structures. SN~1987A's circumstellar rings also contain dust grains and the destruction of these grains by sputtering in encroaching shocks can be studied in real time at infrared wavelengths, providing information of direct relevance to our understanding of dust destruction and survival in the interstellar medium of our own and other galaxies. 

Core-collapse supernovae are the only plausible sources of the relatively large masses of dust found in many very high-redshift galaxies \citep[e.g.][]{Laporte2017}. As discussed in Section 3.1, SN~1987A has proven to be the key object for studies of dust formation in CCSN ejecta - its current ejecta dust mass of $\sim0.5$~M$_\odot$, spatially resolved by ALMA, turns out to be typical of other CCSNe that have been observed at similarly late epochs.

As discussed in Section 3.2, the formation and disappearance of molecules as the ejecta expands and cools has been studied in detail for SN~1987A and has been found to show good agreement with model predictions. Molecules are the building blocks for dust particles and predictions that dust grains should start to condense in CCSN ejecta within a few hundred days after outburst were positively confirmed in the case of SN~1987A. However, the subsequent evolution with time of SN~1987A's total dust mass has been found not to agree at all with model predictions, see Section 3.1.

A prediction of core-collapse supernova models is that the collapse should lead to the formation of a neutron star at the core. SN~1987A offered the opportunity for the earliest detection of a newly formed neutron star. As described in Section~4, searches from X-ray to radio wavelengths during the first 35 years after outburst,  
failed to find evidence for the presence of either a pulsar or a very hot bare neutron star but the search has recently seen success as a result of infrared observations of SN~1987A by the {\em James Webb Space Telescope}.

Several previous reviews have been published on the properties and evolution of SN~1987A and its remnant SNR~1987A (which will both be referred to here as SN~1987A). These include reviews by D.~Arnett et al. in 1989 \citep{Arnett1989}, by R.~McCray in 1993 and 2007 \citep{McCray1993, McCray2007} and by R.~McCray and C.~Fransson in 2016 \citep{McCray2016}. 
Some key results covered by these reviews include:
\begin{itemize}

\item Starting at 07:35 U.T. on February 23rd 1987, 25 anti-neutrinos were 
detected, 12 by the Kamiokande-II detector in Japan \citep{Hirata1987}, 8 
by the IMB detector in the U.S. \citep{Bionta1987} and 5 by the Baksan 
detector in the U.S.S.R. \citep{Alekseev1987}.
They originated from the final core collapse of the star and the likely 
formation of a neutron star, with neutrinos being the sole species that 
could escape from the core carrying the immense amount of energy released 
by the gravitational collapse of the star.

\item
SN~1987A had an unusual optical light curve, with its slow rise from a visual magnitude of 4.3 at discovery to its third magnitude maximum taking three months \citep{Catchpole1987}, attributed to its more compact blue supergiant structure at outburst \citep{Pumo2025}, compared to the much more extended envelopes of M supergiant stars that are usually associated with Type~II supernova events.

\item 
The overall light curve of SN~1987A was dominated by the radioactive decays $^{56}$Ni$\rightarrow^{56}$Co$\rightarrow^{56}$Fe, with an initial $^{56}$Ni mass of 0.069$\pm$0.003~M$_\odot$ \citep{McCray2016} decaying with a half-life of 5.6~days, followed by the decay of $^{56}$Co, with a half-life of 77.3~days, to stable $^{56}$Fe nuclei. Beyond that, the light curve could be accounted for by the slower decay of 0.003~M$_\odot$ of $^{57}$Co to $^{57}$Fe, with a half-life of 272~days. From seven years after outburst the luminosity of SN~1987A's ejecta has been powered by the decay of $^{44}$Ti to $^{44}$Sc, with a half-life of 59.1~years. More recently, the strength of the 4.09~keV X-ray emission line of the $^{44}$Sc decay product has been measured in {\em Chandra} spectra to estimate an initial $^{44}$Ti mass of (1.6$\pm$0.3)$\times10^{-4}$~M$_\odot$ \citep{Giuffrida2025}, in agreement with the initial $^{44}$Ti mass of (1.5$\pm0.3)\times10^{-4}$~M$_\odot$ estimated (a) from {\em NuSTAR} measurements of the 67.87~keV decay line of $^{44}$Ti \citep{Boggs2015} and (b) from an analysis of the energetics of a {\em Hubble Space Telescope (HST)} year~8 optical-UV spectrum of SN~1987A \citep{Jerkstrand2011}.
    
\item
From about day 500, SN~1987A's optical light curve showed a dip below the expected decay luminosity of $^{56}$Co, coinciding with the appearance of excess mid-IR continuum emission and the development of red-blue asymmetries in H$\alpha$ and other emission lines from the ejecta. These phenomena can all be attributed to the formation of dust grains in the ejecta and are discussed in more detail in Section~3.

\end{itemize}

\begin{figure}
\hspace{2.7cm}
\includegraphics[width=0.66\hsize]{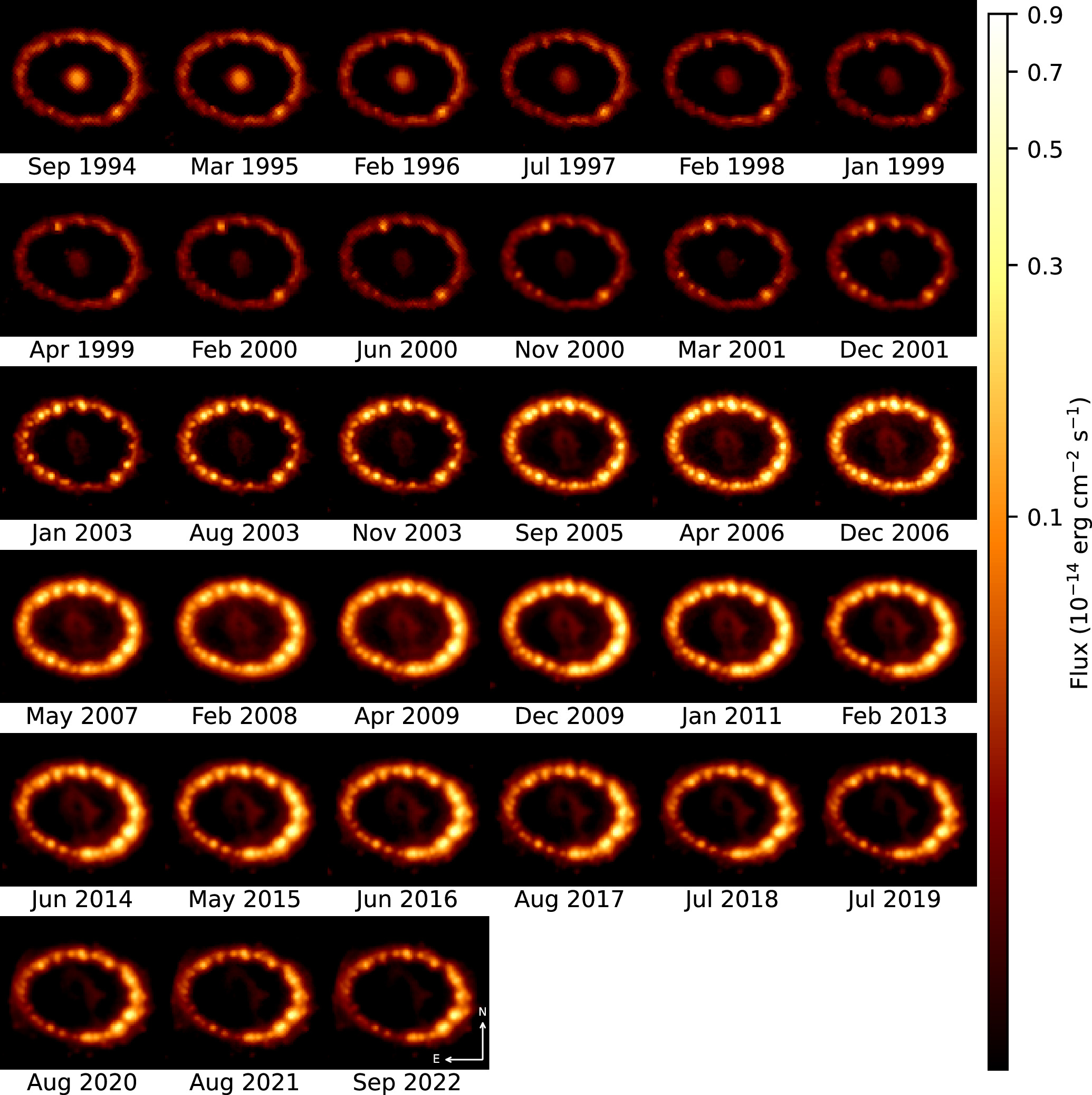}
\caption{Optical R-band {\em HST} images of the equatorial ring taken between 1994 and 2022, showing the appearance, brightening and then fading of 26 hot spots in the ring. The field of view of each image is 2.30$\times$2.15~arcseconds. The spot at approximately 5 o’clock in the early images is a field star. Figure from Tegkelidis et al. 2024 \citep{Tegkelidis2024}.}
\label{Fig:tegkel_fig1}
\end{figure}

\noindent This review will focus on observational and theoretical 
developments during the past ten years in the evolution of SN~1987A, 
although some related earlier works will also be covered. Section~2 covers 
recent studies of the evolution of the gas and dust in the equatorial 
ring, while Section~3 covers recent studies of the gas and dust in the 
ejecta of SN~1987A. Section~4 reviews recent developments in the hunt for 
the elusive remnant neutron star. Finally, Section~5 discusses future 
prospects.

\section{The evolution of the Equatorial Ring and beyond}

Several months after the explosion, three circumstellar rings became visible around SN~1987A, flash-ionized by the pulse of extreme ultraviolet and soft X-ray radiation that accompanied the emergence of the blast wave from the stellar photosphere. The bright inner or equatorial ring (ER) had an angular diameter of 1.6~arcseconds (Figure~1) and is consistent with a circular ring structure inclined at 43 degrees to the line of sight
\citep{Sugerman2005, Tegkelidis2024}.

Since it has proved difficult to find plausible single star scenarios that could produce the observed multiple ring structures seen in Figure~2, a binary stellar merger event during a common-envelope phase has been proposed as having produced the expanding rings \citep{Morris2007, Morris2009} - if so, this might also account for the unusual B supergiant spectral type of the progenitor star. Binary star masses of 14-15~M$_\odot$ and 7-9~M$_\odot$ merging to create the B supergiant progenitor with a mass of $\sim$20~M$_\odot$ have been proposed \citep{Menon2017, Urushibata2018}.

\subsection{The initial flash ionization of the equatorial ring}

The timing of the appearance of the flash-ionized ER following the supernova explosion enabled a radius of 0.65 light-years to be estimated for the ring \citep{Sugerman2005}. 
The brightness of the ER peaked about a year after its first appearance, consistent with the ionization front having reached the outer edge of the ring. The rate of fading of individual knots due to recombination indicated electron densities ranging from $\sim3\times10^4$~cm$^{-3}$ down to $\sim1\times10^3$~cm$^{-3}$ and a total ionized mass in the ring of $\sim0.058$~M$_\odot$ \citep{Mattila2010}.

The expansion velocity of 10.3~km~s$^{-1}$ measured for the ER implied an expansion age of $\sim$20,000~years \citep{Crotts2000}.
The fainter outer two rings have angular diameters about twice as large as that of the ER but their 2.5 times larger expansion velocities are consistent with their ejection having taken place at about the same time as that of the ER. 

\subsection{Interaction of the ejecta with the equatorial ring}

The outermost and fastest moving layers of the ejecta began to impact the ER in 1995 \citep{Sonneborn1998}, manifested by rapidly brightening hot spots, with the first one appearing in the northeastern part of the ring. By 2003 the whole ring contained 26 hot spots \citep{Tegkelidis2024} -- see Fig.~\ref{Fig:tegkel_fig1}. The hot spots have been interpreted as resulting from shocks propagating into dense clumps within the ER \citep{Chevalier2017}. Rayleigh-Taylor instabilities\footnote{The Rayleigh-Taylor instability occurs at the interface between two fluids of different densities when a less dense fluid moves into a denser fluid, causing plumes of material to inter-penetrate.} in the ring prior to the supernova have been suggested as the cause of the clumps (e.g. \citep{Orlando2015}), although the hydrodynamic Crow instability\footnote{The Crow instability is where a pair of counter-rotating vortices act upon each other to amplify small sinusoidal distortions in their vortex shapes. Eventually, the vortex amplitudes reach a critical value and reconnect, forming vortex rings.}, prior to the supernova, has also been proposed as accounting for the sizes and numbers of the clumps \citep{Wadas2024}.

\begin{figure}
\centering
\includegraphics[width=0.58\hsize]{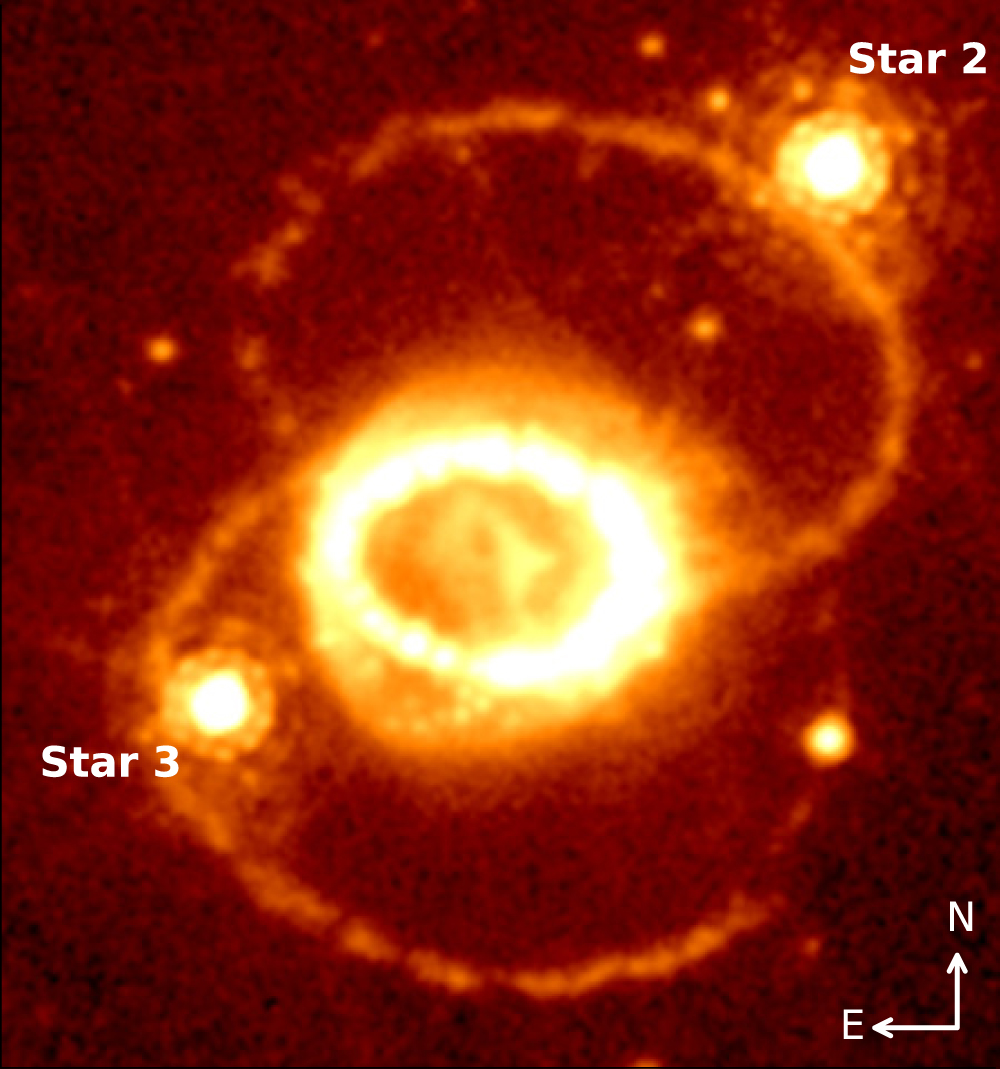}
\caption{A 2018 (day 11,500) {\em HST}/WFC3 F657N image of the triple ring system around SN~1987A and the fainter emission just outside the bright equatorial ring (ER). The F657N filter has a FWHM of 121~\AA\ and encompasses the H$\alpha$ and [N~{\sc ii}] 6584,6548~\AA\ lines. Inside the ER, the faint H$\alpha$ emission from the inner ejecta is produced by external irradiation by X-rays from the shocked ER. The X-ray emission is stronger on the western side of the ER \citep{Frank2016}. Figure from Larsson et al. 2019 \citep{Larsson2019b}.}
\label{Fig:larsson19_fig1}
\end{figure}

From X-ray observations, blast wave velocities v$_{blast} \sim 4100-6700$~km~s$^{-1}$, in the ambient low-density flash-ionized gas near the ER, have been estimated \citep{Borkowski1997, Ravi2024, Tegkelidis2024}. On encountering
much higher density material the shock will propagate at lower velocities. v$_{clump}$, the shock velocity in the clump, is proportional to v$_{blast}$(n$_{ambient}$/n$_{clump}$)$^{1/2}$, where n$_{ambient}$ and n$_{clump}$ are the respective volume number densities in the ambient and clumped gas. In order for shocked gas in the ER to become visible in optical emission lines, its cooling time to reach temperatures $\leq2\times10^4$~K must be less than 20-30~years. From this, Tegkelidis et al. \citep{Tegkelidis2024} have estimated that the observed clump emission regions must have pre-shock densities n$_{clump}$ of at least 3$\times10^4$~cm$^{-3}$ and shock velocities v$_{clump} \leq 500$~km~s$^{-1}$, with each hot spot comprised of dense substructures embedded in less dense gas. Lower clump region densities would lead to such regions not yet having become optically visible due to the slower cooling rates.
The measured widths of optical emission lines from ER hot spots indicate velocities of 200-500~km~s$^{-1}$ in the post-shock gas \citep{Groningsson2008}, consistent with the above considerations.

The impact on the ER of the blast wave at the leading edge of the ejecta has caused the ER to expand, with best-fit expansion velocities of 680$\pm$50~km~s$^{-1}$ and 690$\pm$190~km~s$^{-1}$ measured at optical and near-IR wavelengths, respectively \citep{Larsson2019b, Kangas2023}, from the ER's size increases. The increasing brightness of the ER at optical and near-infrared wavelengths, dominated by line emission from the hot spots, plateaued in 2009 (day $\sim$8000) and then began to decline \citep{Fransson2015, Kangas2023} as the blast wave passed through the ER and began to interact with less dense matter outside the ring.

Beyond the ER, a dozen fainter spots, mostly in the southeast, and a rim of diffuse H$\alpha$ emission appeared after day $\sim$9500 (see Fig.~\ref{Fig:larsson19_fig1}) and were interpreted as fast ejecta interacting with high latitude material that extends from the ER toward the outer rings. Larsson et al. \citep{Larsson2019b} noted that with the then current rate of decline of the ER's R-band flux, its flux should reach zero by 2035.

\subsection{JWST observations of the equatorial ring}

The launch of the {\em James Webb Space Telescope (JWST)} on December 25th 2021 enabled observations to be made of SN~1987A with unprecedented sensitivity and angular resolution over its 1-28-$\mu$m wavelength range. Results based on imaging using the Near-Infrared Camera (NIRCam) \citep{Arendt2023, Matsuura2024} and the Mid-Infrared Instrument (MIRI) \citep{Bouchet2024} have been published, together with spectroscopic/imaging results using the integral field units (IFUs) on-board the Near-Infrared Spectrograph (NIRSpec) and the MIRI-Medium Resolution Spectrometer (MRS) \citep{Larsson2023, Jones2023b, Fransson2024}. 

\begin{figure}
\centering
\includegraphics[width=0.68\hsize]{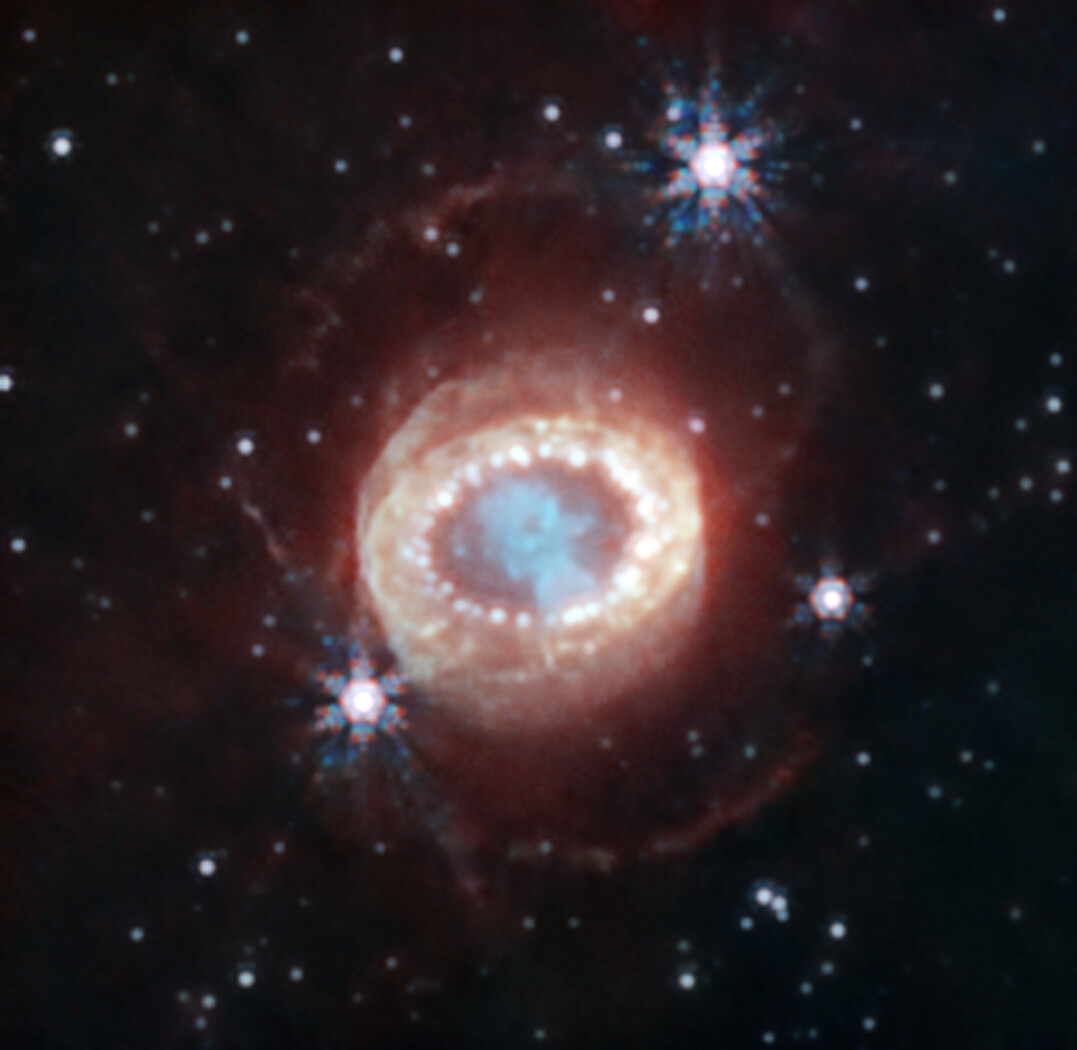}
\caption{A 2022 (day 12,974) {\em JWST}-NIRCam five-filter composite image of SN~1987A \citep{Matsuura2024}. The cyan colour of the inner ejecta corresponds to strong emission in the F164N filter, which encompasses [Fe~{\sc ii}] 1.6436~$\mu$m and [Si~{\sc i}] 1.6455~$\mu$m. 
Image credit: NASA, ESA, CSA, M. Matsuura et al.}
\label{Fig:87A_nircam}
\end{figure}

\begin{figure}
\centering
\includegraphics[width=1.0\hsize]{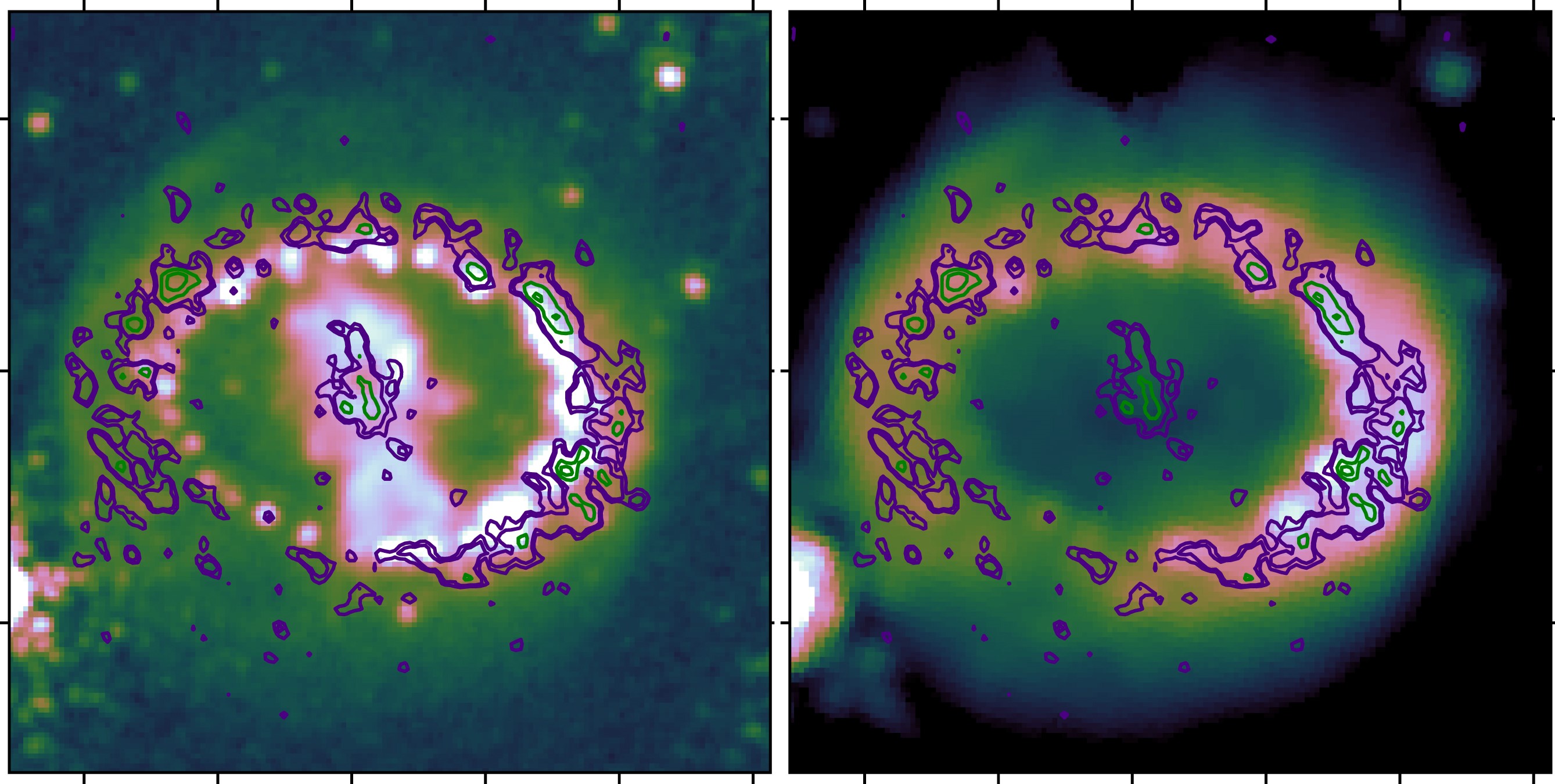}
\caption{{\em JWST} NIRCam F164N image (left) and F356W image (right), with ALMA 315~GHz continuum emission shown as contour lines. The ALMA emission is dominated by nonthermal synchrotron radiation from the ER and dust emission from the ejecta. The ER hotspots and the outer spots detected in the F356W image generally have corresponding spots in the ALMA 315 GHz synchrotron emission.
From Matsuura et al. (2024 \citep{Matsuura2024})}
\label{Fig:Matsuura24_fig8}
\end{figure}

Fig.~\ref{Fig:87A_nircam} shows a NIRCam composite image of SN~1987A taken with five filters covering the 1-5-$\mu$m range \citep{Matsuura2024}, while Fig.~\ref{Fig:Matsuura24_fig8}, also from Matsuura et al. (2024 \citep{Matsuura2024}), compares a NIRCam F164N (1.64~$\mu$m) image, dominated by [Fe~{\sc ii}] and [Si~{\sc i]} line emission, with a F356W (3.6~$\mu$m) image, dominated by synchrotron emission. Contours from their 2021 Atacama Large Millimetre Array (ALMA) 315~GHz map, also  dominated by nonthermal synchrotron emission in the ER, are shown as well. The ALMA contours fall outside the hotspots seen in the F164N image but coincide with the hotspots seen in the F356 image, consistent with the blast wave having progressed outside the ER and into the diffuse gas beyond, where it produces the synchrotron radiation that dominates both the NIRCam F356 and ALMA 325~GHz continuum. The outer radius of the nonthermal radio and X-ray emission has also moved outwards with time \citep{Cendes2018, Frank2016}.

\begin{figure}
\centering
\includegraphics[width=1.0\hsize]{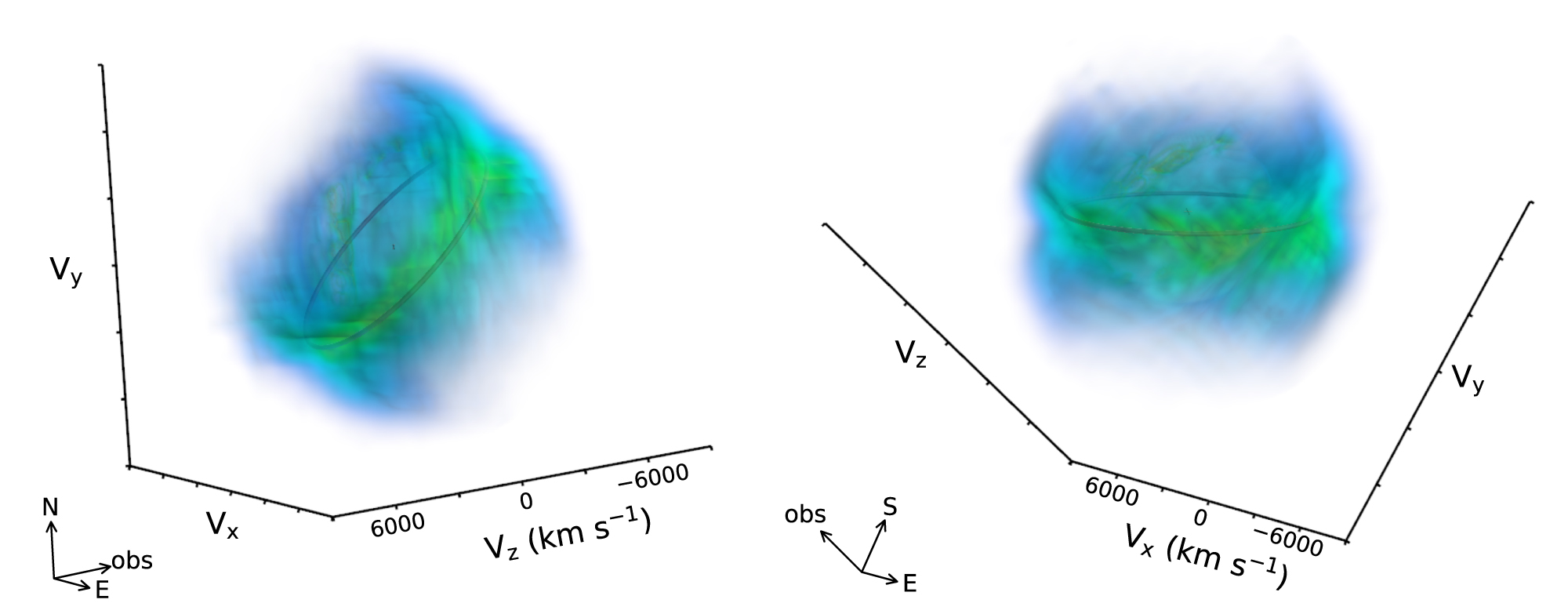}
\caption{{\em JWST} NIRSpec volume rendering of the day~12,927 He~{\sc i} 
1.083-$\mu$m emission from SN~1987A. The emission from the inner ejecta velocities ($<$4000~km~s$^{-1}$) has been omitted. The grey circle shows the position of the ER. From Larsson et al. (2023 \citep{Larsson2023}).
}
\label{Fig:Larsson23_10830}
\end{figure}

NIRSpec IFU spectra of SN~1987A, obtained on day 12,927 in 2022 with a 
spectral resolving power R$\sim1000$ and a field of view of 
3.3$\times3.8$~arcsec, enabled 3D emissivity maps to be constructed for 
multiple lines, assuming free expansion since the explosion 
\citep{Larsson2023}. Ejecta material which has reached the ER's semi-major 
axis of 0.82~arcsec must have had an expansion velocity of 
5400~km~s$^{-1}$ for an LMC distance of 49.6~kpc \citep{Pietrzynski2019}, 
while emission at higher velocities is expected to originate at higher 
latitudes above and below the ER plane. Fig.~\ref{Fig:Larsson23_10830} 
displays the 3D image constructed for the He~{\sc i} 1.083-$\mu$m line. It 
shows the emission to originate from a surface that extends from the inner 
edge of the ER to higher velocities on both sides of it, forming a 
bubble-like structure at higher latitudes above and below the ER. The 
pointing of the high latitude emission towards the outer rings may 
indicate that the material is part of a structure that connects the three 
rings \citep{Larsson2023}. A similarly constructed 3D image constructed 
for the [Fe~{\sc i}] 1.443-$\mu$m line from the inner ejecta shows an 
elongated morphology with the brightest emission concentrated in a 
blue-shifted clump in the north and a red-shifted clump in the south, with 
peak 3D space velocities of $\sim$2300 and $\sim$2200~km~s$^{-1}$, 
respectively \citep{Larsson2023}.

MIRI-MRS IFU observations of SN~1987A were also obtained on day 12,927 
\citep{Jones2023b}, with spectral resolutions of $\sim$80-160~km~s$^{-1}$ depending on wavelength. From a comparison of emission line full width at half maximum (FWHM) values and radial velocities, it was found that all lines originating from singly ionized species showed broad line-widths (FWHMs $\sim$ 200-300~km~s$^{-1}$) along with radial velocity variations around the ER, whereas all lines from doubly ionized or higher species (e.g. S$^{2+}$ up to Ne$^{5+}$) showed narrower lines (FWHM $\sim$ 100-160~km~s$^{-1}$) with no radial velocity variations seen at different positions on the ER. This is consistent with all of the lines from singly ionized species originating from the dense expanding ER, while all the lines from more highly ionized species originate from more extended material. The latter component was attributed to very extended circumstellar gas which had been flash-ionized by the pulse associated with the supernova event and whose low density has prevented it from recombining yet (given that timescales for recombination scale inversely with ambient electron density).

VLT/UVES 6~km~s$^{-1}$ resolution observations obtained in 2002 of northern and southern parts of the ER  \citep{Groningsson_vel2008}
had found FWHM line widths of 10-29~km~s$^{-1}$ for a narrow-line component, consistent with the narrow lines seen by the MIRI-MRS being
spectrally unresolved by the MRS. Electron densities of 1500-5000~cm$^{-3}$ were derived from emission line diagnostic ratios measured in the UVES spectrum \citep{Groningsson_vel2008}, in agreement with the electron densities of 700-4300~cm$^{-3}$ that were derived for the MRS narrow line component from measured [Ne~{\sc v}] 14.32-$\mu$m/24.32-$\mu$m
line ratios \citep{Jones2023b}.

\subsection{Hydrodynamical/MHD models for the interaction between the ejecta and the equatorial ring}

\begin{figure}
\centering
\includegraphics[width=0.7\hsize]{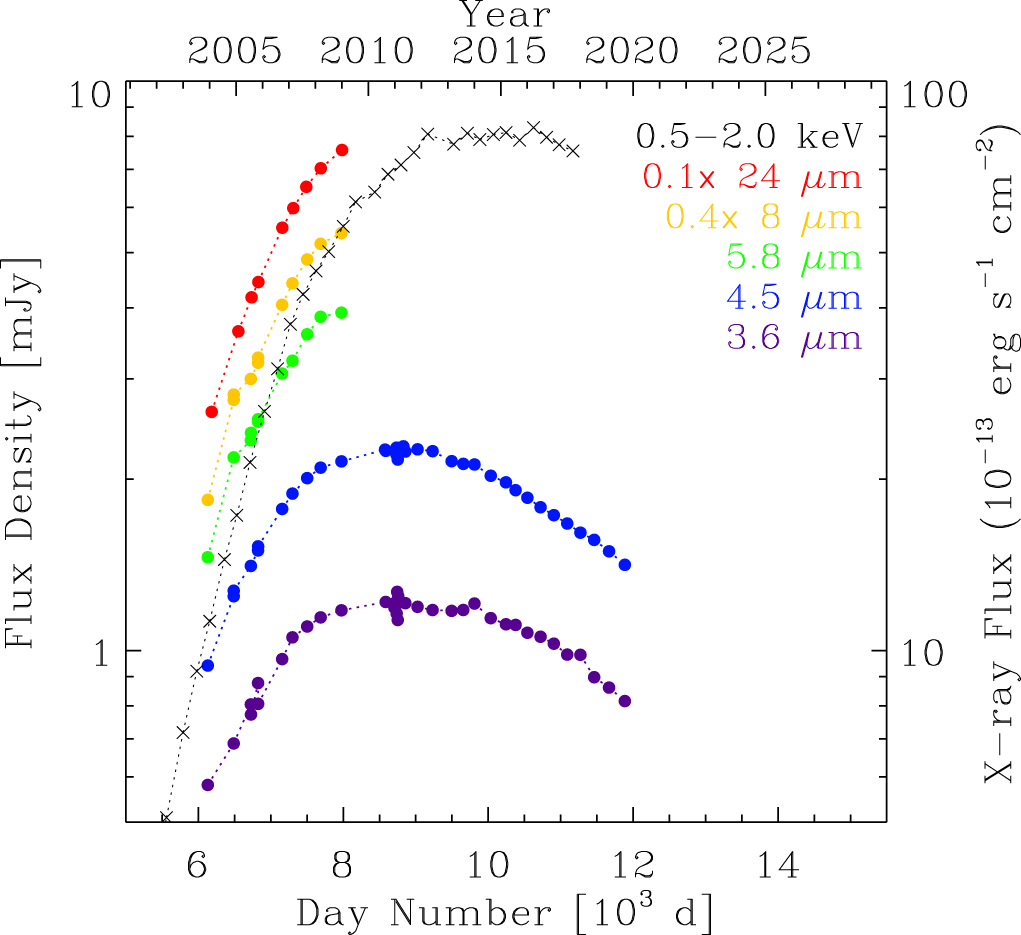}
\caption{SN 1987A light curves through the entire {\em Spitzer} mission, with the {\em Chandra} soft X-ray light curve (0.5–2 keV, \citep{Frank2016}) shown for comparison (black x symbols). From Arendt et al. 2020 \citep{Arendt2020}.
}
\label{Fig:arendt20_fig1}
\end{figure}

Orlando et al. \citep{Orlando2015, Orlando2019} have modelled the 3D hydrodynamic interaction between the expanding remnant of SN~1987A and the surrounding circumstellar material (CSM), including the equatorial ring. To do this they first modelled in 1D the early post-explosion evolution of the SN during the initial 24 hours and then mapped the output of these simulations into 3D in order to model the remnant's interaction with the inhomogeneous CSM. For early epochs ($<$15 years) they found that the shape of the X-ray light curve was determined mainly by the structure of the outer ejecta, with the best fit to the X-ray observations being found with a post-explosion radial density profile approximated by a power law index of n = -8. For years 15-32, where the outer ejecta interact with the ER, soft X-ray emission is dominated by the shocked clumps while the hard X-ray emission originates from the smooth interclump regions \citep{Orlando2015}. For a third phase, beyond year 32, the forward shock (blast wave) was predicted to propagate beyond the ER, with a reverse shock predicted to move back into the expanding ejecta, leading to very strong X-ray emission from the shocked ejecta by 2021 - this prediction has subsequently been confirmed by {\em XMM-Newton} X-ray observations \citep{Sun2025}.

The 3D hydrodynamical models described above were subsequently extended to 3D magnetohydrodynamical simulations by considering ambient magnetic field strengths between 1 and 100~$\mu$G \citep{Orlando2019}. It was found that, compared to the no-field case, the presence in the ER of magnetic fields in this range had the effect of reducing the erosion and fragmentation of the ring, so that it survived the passage of the blast wave during the 40 years covered by the simulations.

\subsection{The dust in the equatorial ring}

Using {\em Spitzer Space Telescope} mid-infrared (MIR) spectroscopy and photometry and ground-based 8-m telescope MIR imaging obtained between days 6067-6526 (years 16.6-17.9), Bouchet et al. \citep{Bouchet2006} showed that the ER contained warm (T$\sim$170~K) silicate dust exhibiting broad emission peaks at 10 and 20~$\mu$m.
Dwek et al. \citep{Dwek2008, Dwek2010} presented MIR {\em Spitzer} data obtained over a five year period up to day 7955 and concluded that the silicate dust was being collisionally heated to T$\sim$180~K by gas particles in the ER's shocked X-ray emitting region, with a gas temperature and density of $\sim5\times10^6$~K and $\sim3\times10^{4}$~cm$^{-3}$, respectively. They 
showed that the ER's value of 2.5 for the ratio of total IR flux to total X-ray flux implied that gas-grain collisions dominate the cooling of the shocked X-ray emitting gas. The constancy of this ratio over the five year period was interpreted by them as implying that very little grain destruction or gas cooling had occurred during this time period. This was despite the fact that the total IR flux had increased by a factor of five during the time in question, responding to the blast wave transitioning from free expansion to a Sedov phase\footnote{The Sedov phase of evolution of a supernova remnant follows the free expansion phase
and begins when part or all of the remnant has swept up a mass of circumstellar or interstellar gas similar to its own.}, in which there is a significant mass of swept-up gas as it propagates into the dense ER \citep{Dwek2010}.

Arendt et al. \citep{Arendt2016, Arendt2020} monitored the  3.6-24-$\mu$m MIR fluxes of SN~1987A's ER from shortly after the launch of {\em Spitzer} in 2003 up to the exhaustion of its liquid helium cryogens in 2009, after which 3.6- and 4.5-$\mu$m photometry continued to be acquired until the termination of the warm {\em Spitzer} mission in 2020. Fig.~\ref{Fig:arendt20_fig1} compares the time evolution of the MIR fluxes from the ER with its 0.5-2~keV soft X-ray light curve. The 8- and 24-$\mu$m fluxes track the soft X-ray flux up to the end of the cold {\em Spitzer} mission in 2009, after which the X-ray flux began to plateau and then decline, as the leading shock traversed and then exited the ER. However, the 2003-2019 3.6- and 4.5-$\mu$m fluxes began to plateau and then decline earlier than the soft X-ray flux, which was attributed by them to the smaller dust particles, which were considered to be responsible for the 3.6-4.5-$\mu$m hotter dust emission, being more easily destroyed by sputtering than the larger grains responsible for the 10- and 20-$\mu$m dust emission \citep{Arendt2016, Arendt2020}.\footnote{For grain sputtering by gas particles, the rate of decrease of the grain radius $da/dt$ is independent of the grain radius $a$, so smaller grains will be destroyed more quickly than larger grains.}

\begin{figure*}
\centering
\includegraphics[width=0.50\hsize]{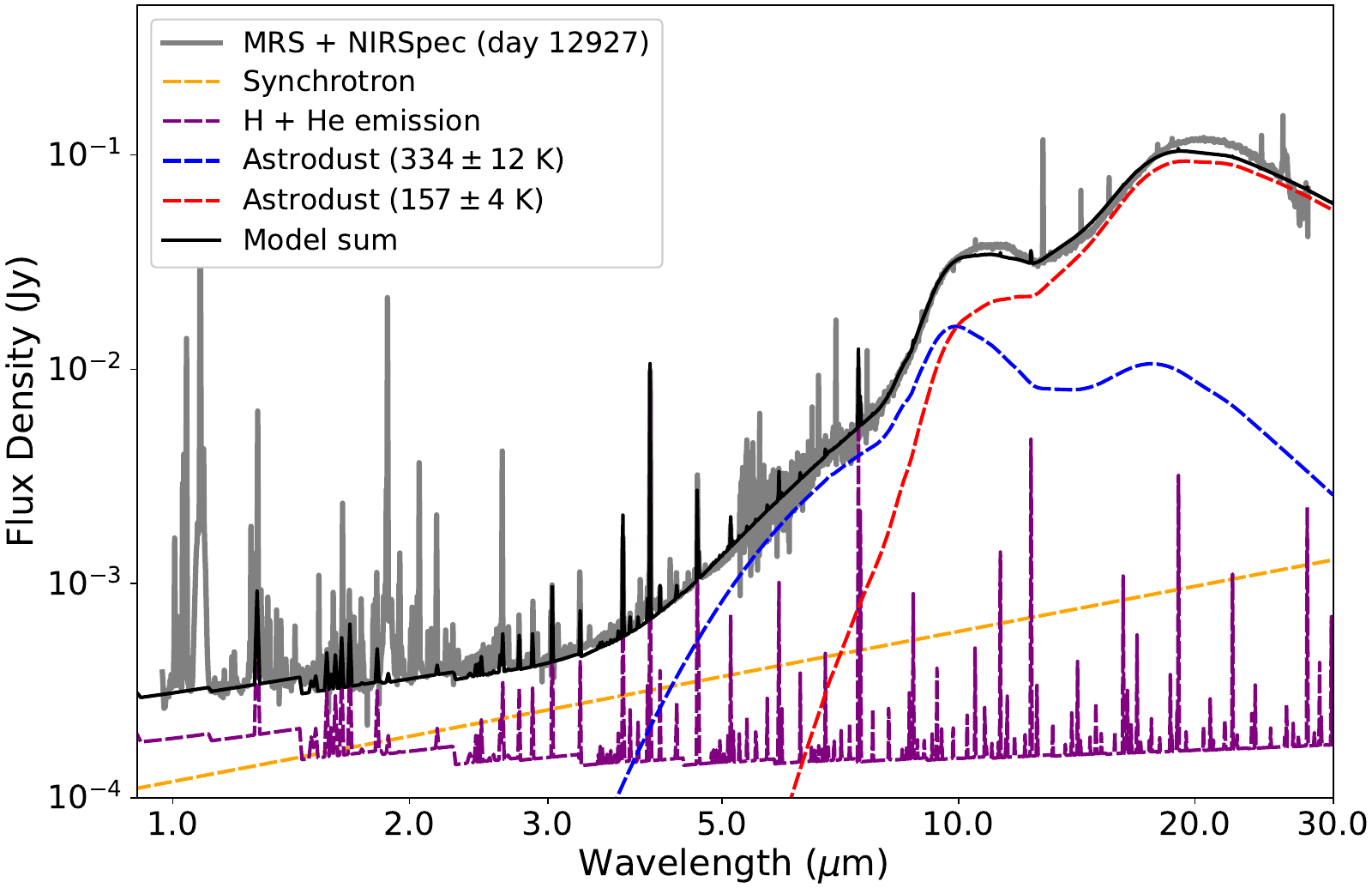}
\includegraphics[width=0.48\hsize]{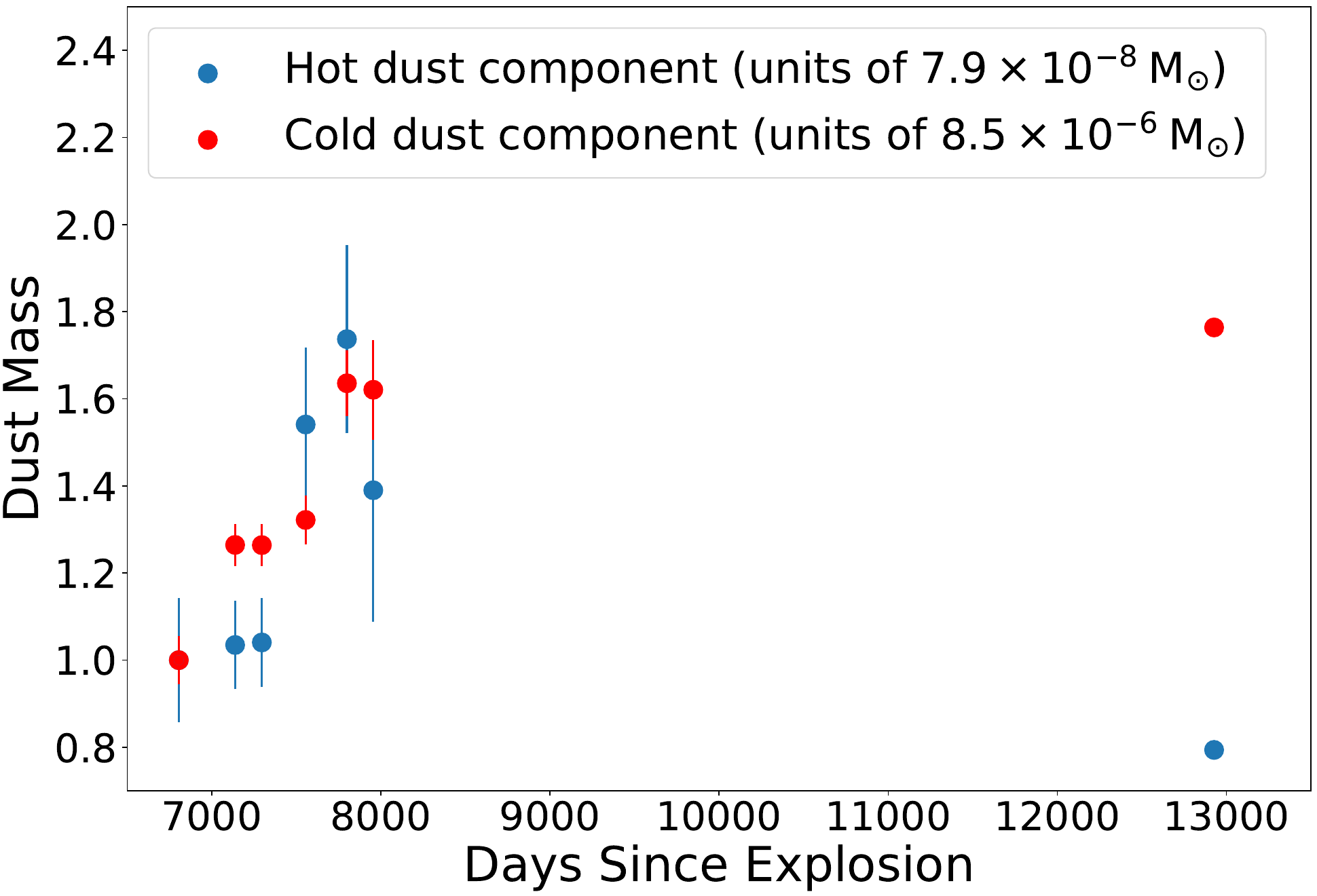}
\caption{Left panel: A sample fit to the day 12,927 {\em JWST} MRS+NIRspec 
spectrum of the ER using two-temperature {\em astrodust} \citep{Hensley2023} components.
Right panel: Best fit derived dust masses using a two-temperature {\em 
astrodust} model for each epoch's spectrum, fitted over the 5-30-$\mu$m spectral range ({\em Spitzer} IRS up to day 7955, MIRI-MRS on day 12,927). The error bars represent 3$\sigma$ uncertainties. Figures from Jones et al. 2023 \citep{Jones2023b}.
} 
\label{Fig:dustmass_evol}
\end{figure*}

The {\em JWST} MRS+NIRSpec spectrum of the ER \citep{Jones2023b} provided much higher spatial and spectral resolution than the prior {\em Spitzer}-IRS spectra. Dust emission fits to both the MRS and IRS spectra, using the same Hensley \& Draine \textit{astrodust} optical constants for all epochs, found similar dust temperatures at each epoch for both the `hot' dust component (T$\sim382\pm12$~K) and the `cold' dust component (T$\sim162\pm2$~K).
The mass of ($6.3\pm0.4)\times10^{-8}$~M$_\odot$ derived for the hot dust 
component in the day 12,927 {\em JWST} spectrum (see Fig.~\ref{Fig:dustmass_evol}-left) showed a decrease by about a factor of two compared to the hot dust component masses derived from the last two {\em Spitzer} spectra, on days 7799 and 7955, consistent with ongoing destruction of the smallest grains in the ER (see Fig.~\ref{Fig:dustmass_evol}-right). In contrast, the cold dust component mass of (1.50$\pm0.03)\times10^{-5}$~M$_\odot$ 
deduced from the fit to the day 12,297 MRS spectrum was unchanged within 
the uncertainties from the cold dust component masses derived from the last two epochs of {\em Spitzer}-IRS spectra (Fig.~\ref{Fig:dustmass_evol}-right), consistent with negligible destruction during that interval of the larger grains responsible for the cold component. 

Combining the above estimate of $1.5\times10^{-5}$~M$_\odot$ for the mass of dust in the ER with the gas mass of 0.058~M$_\odot$ derived for the ER when it was flash-ionized \citep{Mattila2010}, implies a dust-to-gas mass ratio in the ER of 1/3870, which is a factor of 11 lower than current estimates for the dust-to-gas mass ratio in the LMC interstellar medium \citep{Clark2023}\footnote{Clark et al. \citep{Clark2023} derived a dust-to-hydrogen mass ratio of 1/245 for the LMC ISM. Combined with an LMC He/H number ratio of 0.094 \citep{Tsamis2003}, this implies a dust-to-gas mass ratio of 1/338 for the LMC.}.

The fact that {\em astrodust} optical constants, which had been devised to fit the optical and infrared spectra of Milky Way interstellar dust \citep{Hensley2023}, provide a reasonable fit to the IR spectrum of the circumstellar dust in the ER may seem surprising. However, no predictions exist for what types of dust should form during a mass ejection resulting from a massive binary star common envelope phase followed by a merger event.

\subsection{The evolution of the equatorial ring and beyond: summary}

The ability to obtain spatially resolved observations of the SN~1987A ring system at X-ray through to radio wavelengths 
has allowed us to study the onset and time evolution of shocks as they encounter regions having vastly different densities. The forward shock (blast wave), travelling at velocities of 5100-6700~km~s$^{-1}$ through the low-density ambient gas outside the ring system, produces hard X-ray emission from the resulting shocked gas. The cooling time of this very hot gas is long and so the shock remains adiabatic in such regions. In the ER hot spots, densities in excess of 10$^4$~cm$^{-3}$, with shock velocities of $\leq 500$~km~s$^{-1}$, led to soft X-ray emission in the immediate post-shock region, while the subsequent rapid cooling of the dense post-shock gas led to the establishment of isothermal conditions, accompanied by strong UV, optical and IR line emission. We now know that early interactions between outflowing supernova ejectae and pre-existing circumstellar material is common around most types of core collapse supernovae. Not only can we study such interactions in greater detail for SN~1987A than for much more distant extragalactic SNe but, uniquely, from our early observations of SN~1987A we know the geometries and the physical conditions in its circumstellar structures {\it before} the supernova blast wave reached them and shocks were formed. 

The high velocity leading edge of the blast wave has already passed through the ER and is currently illuminating lower density structures beyond the bright ring (Figures 2 and 3). The asymmetric main body of the ejecta, expanding at just $\sim$1600~km~s$^{-1}$ and visible in those figures due to its external irradiation by UV and X-ray photons from regions already shocked by the fast blast wave, is only now beginning to reach the ER. In the coming years the dense ejecta material will overrun the ER, destroying not just the dust particles located there but also the dense clumps of gas that make up the hot spots. SN~1987A will provide us with a ringside view of time-dependent processes that we know to be commonly encountered in interstellar and circumstellar regions but which cannot be studied anywhere else at the same level of detail. 

\section{The SN 1987A ejecta}

\subsection{Dust formation in the ejecta of SN~1987A}

\subsubsection{The first four years after outburst}

Starting 400-500 days after outburst, strong evidence emerged for dust having formed in SN~1987A's ejecta. This included the development of a 3-20-$\mu$m mid-infrared continuum emission excess which greatly exceeded that expected from the extrapolated optical continuum or from atomic continuum emission processes \citep[e.g.][]{Bouchet1991, Roche1993, Wooden1993}. In addition, red-blue optical line profile asymmetries developed that were attributed to the preferential absorption by dust within the ejecta of redshifted emission line photons traversing from the far side of the ejecta to the observer \cite{Lucy1989}. Those authors modelled the observed line profile of the [O~{\sc i}] 6300,6363~\AA\ doublet at day 775 after outburst to derive a total ejecta dust mass of 3$\times10^{-4}$~M$_\odot$ at that epoch, for the case of silicate grains. At the same epoch, a fit to SN~1987A's mid-IR continuum emission, observed with NASA's {\em Kuiper Airborne Observatory (KAO)}
by \cite{Wooden1993} using a clumped dust distribution  yielded a similar dust mass, 5$\times10^{-4}$~M$_\odot$. However, this was a factor of 20 larger than the dust mass obtained from the day~615 mid-IR excess emission, indicating increasingly efficient dust condensation between days 615 and 775 \citep{Wooden1993}.

A trend of a decreasing dust temperature with time was also apparent from the mid-IR continuum fitting, with a mean dust temperature of $\sim$640~K found on day 415, $\sim$400~K on day 615 and $\sim$300~K on day 775 \citep{Wooden1993}. This temperature decline can be attributed (a) to the decreasing radiation density in the ejecta as it continues to expand, and (b) to a steady decline in the luminosity of radioactive $^{56}$Co which, with a half-life of 77 days, becomes the main heating source in the ejecta of core-collapse SNe after a few tens of days. $^{56}$Co is ultimately superseded as the main heating source at very late epochs by $^{44}$Ti, whose half-life is 62 years. The declining dust temperature progressively shifts the peak of its reradiated energy to longer and less easily detected infrared wavelengths. 

The last mid-IR detection of SN~1987A's ejecta dust emission took place using the {\em KAO} on day 1153, at wavelengths between 16-29-$\mu$m \citep{Dwek1992}. When combined with ground-based 10-$\mu$m photometry of SN~1987A obtained on day~1100 \citep{Bouchet1993},
an analysis of the mid-IR spectral energy distribution (SED) yielded an ejecta dust mass of 3$\times10^{-3}$~M$_\odot$ on day 1153 \citep{Wesson2015}. During their respective 1995-1998 and 2003-2009 cold missions, the 0.6-m {\em Infrared Space Obsevatory} and the 0.8-m {\em Spitzer Space Telescope} did not have sufficient angular resolution at mid-IR wavelengths  to spatially resolve any inner ejecta emission from the very bright mid-IR emission from the ER. However, ground-based observations of SN~1987A in 2003 (on day 6067) with the 8.1-m Gemini South telescope \citep{Bouchet2004} did detect the inner ejecta at 10~$\mu$m, with a flux of 0.31$\pm$0.10~mJy but, with only a detection at a single wavelength, an ejecta dust mass estimate from SED fitting is not possible for that epoch.

\subsubsection{The detection of cool dust emission from the ejecta 23 years after outburst}

In 2010, as part of the HERITAGE far-infrared survey of the Large Magellanic Cloud \citep{Meixner2010}, on days 8467 and 8564 the 3.5m {\em Herschel Space Observatory} detected an unresolved source at 100, 160, 250 and 350~$\mu$m that was coincident with the position of SN~1987A. The analysis of the energy distribution by Matsuura et al. \citep{Matsuura2011} yielded a dust mass of 0.4-0.7~M$_\odot$ emitting at a temperature of between 17 and 23~K. This result carried the implication that the dust mass had continued to grow after year 3, increasing by a factor of 100 by year 23 after outburst. The magnitude of the year 23 dust mass also implied that the majority of the condensable heavy elements released by the supernova event had been incorporated into dust grains by that epoch \citep{Matsuura2011}.

\begin{figure}
\centering
\includegraphics[width=0.82\hsize]{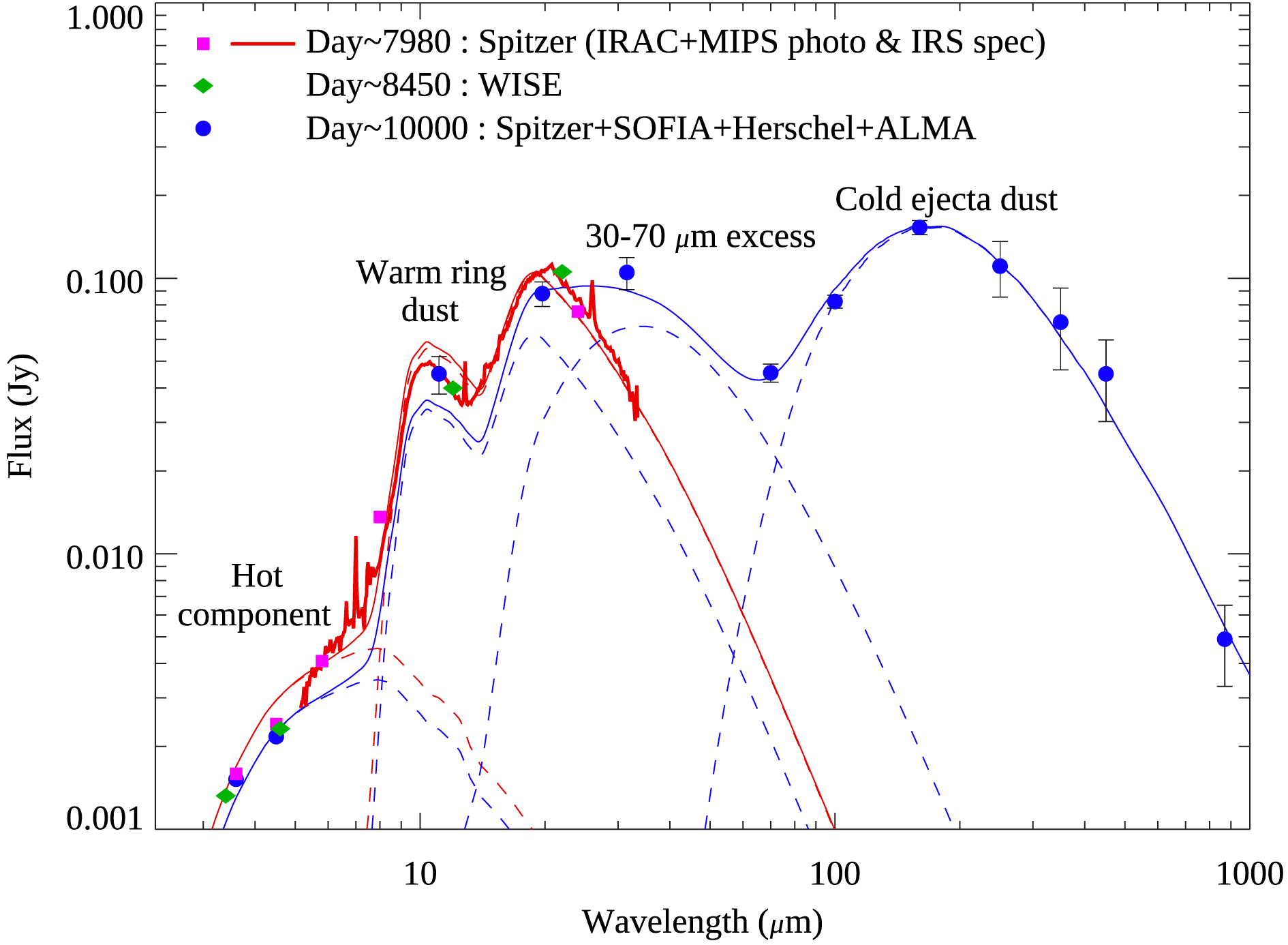}
\caption{The near- to far-infrared SED of SN~1987A at three different epochs, with fitted dust models. The pink squares and thick red line respectively show the {\em Spitzer} IRAC and MIPS photometric data and the IRS spectrum at day 7980, with model fits as thin red lines. 
WISE mid-infrared fluxes at day
8450 are plotted as green diamonds. The blue circles show far-IR/submm flux measurements around day 10,000. The solid blue lines show dust component fits to the SED at day 10,000, with individual components plotted as dashed blue lines. From Cigan et al. 2019 \citep{Cigan2019}.}
\label{Fig:sofia_sed}
\end{figure}

Although the angular resolution of {\em Herschel's} photometers ranged from just 6.8~arcsec at 100~$\mu$m to 25~arcsec at 350~$\mu$m, any doubts about the association between SN~1987A and the unresolved source detected by {\em Herschel} were removed by its 2012 detection \citep{Indebetouw2014} as a point source by ALMA at a wavelength of 442~$\mu$m, where the angular resolution of the array was 0.25$\times$0.33~arscec. This ruled out the 1.8~arcsec diameter ER as the source of the far-IR/submm dust emission, conclusively identifying the inner ejecta as the dust emission source. The inner ejecta was also detected at 870~$\mu$m by ALMA \citep{Indebetouw2014}, while ground-based APEX observations at 350~$\mu$m \citep{Lakicevic2012} provided a flux measurement that agreed with the {\em Herschel} value. Further observations were obtained with {\em Herschel} in 2012, on days 9090 and 9122, again yielding flux measurements with the PACS and SPIRE imaging photometers at 100, 160, 250 and 350~$\mu$m while providing a first detection (13$\sigma$) at 70~$\mu$m \citep{Matsuura2015}. 
Analyses of the combined 2010 and 2012 {\em Herschel} and ALMA flux measurements between 100 and 870~$\mu$m 
yielded an ejecta dust mass of between 0.5 and 0.8~M$_\odot$ at years 23 to 25 \citep{Matsuura2015, Wesson2015}.


The 2012 {\em Herschel}-PACS 70-$\mu$m flux showed an excess above the level expected from the Wien tail of the cold ejecta dust SED, while the 30-$\mu$m flux measured by the {\em SOFIA} airborne observatory on day 10732 (year 29.4) \citep{Matsuura2019} showed an excess above the level expected from the Rayleigh-Jeans portion of the warm ER dust's SED - see Fig.~\ref{Fig:sofia_sed}. The 30-70-$\mu$m excess has been interpreted \citep{Matsuura2019} as being caused by ejecta dust being heated by X-rays from the surrounding interacting ER or, alternatively, as due to dust reformation in downstream post-shock cooling regions where the forward shock is interacting with the ER.

Further high angular resolution ALMA measurements of SN~1987A's ejecta dust continuum fluxes were obtained in year 28.5 at eight frequencies between 225 and 679~GHz (442-1330~$\mu$m) \citep{Cigan2019}, see Fig.~\ref{Fig:sofia_sed}. 
Fits to the SED made using amorphous carbon or silicate dust emission models both give reasonable fits to the data for total dust masses of 0.2-0.4~M$_\odot$ \citep{Cigan2019}. An absolute maximum of 0.7~M$_\odot$ of dust can be present for a grain mixture based on predicted nucleosynthetic yields.

\subsubsection{The timescale for the growth of the dust mass formed in SN~1987A's ejecta}

\begin{figure}
\centering
\includegraphics[width=0.75\hsize]{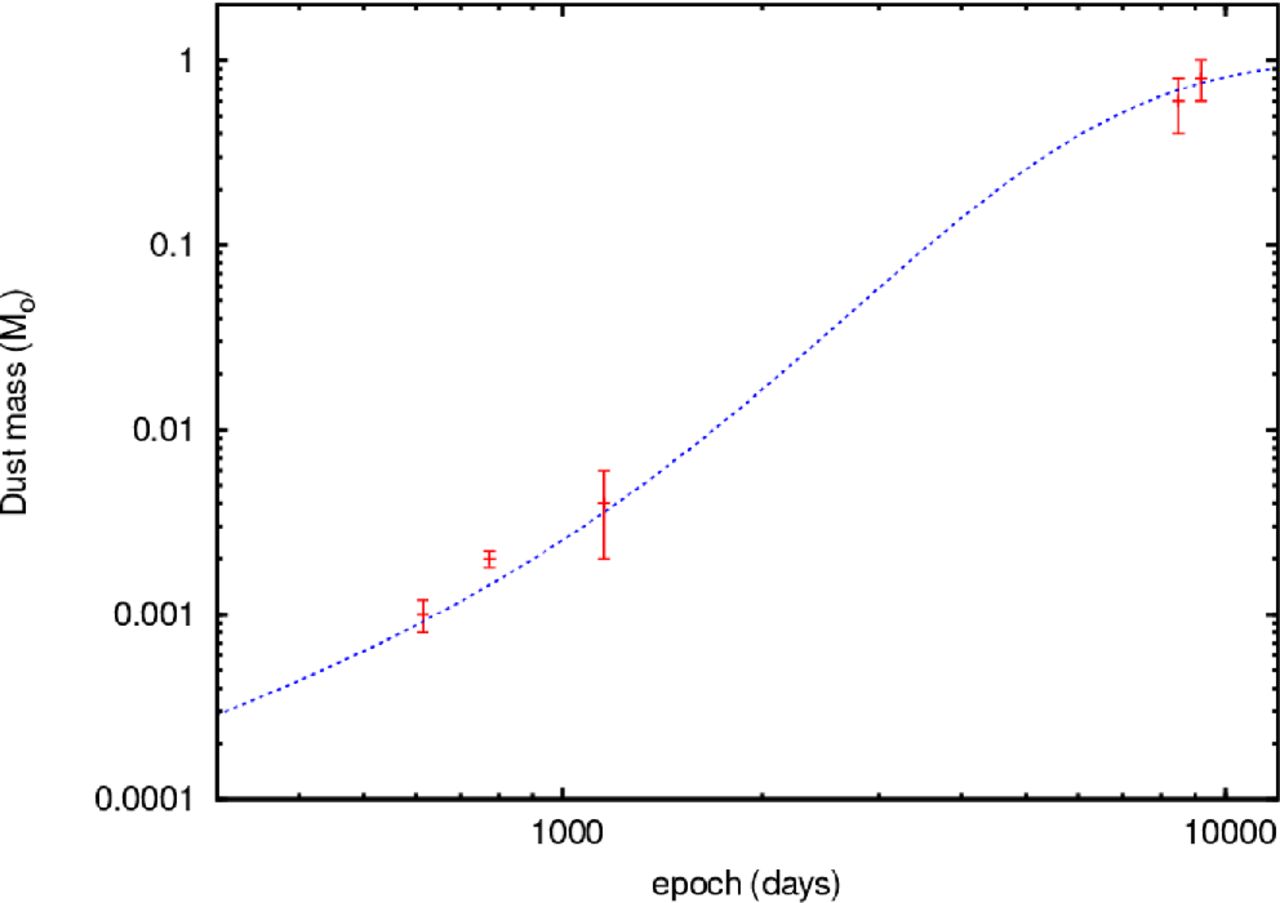}
\caption{The dust mass in the remnant of SN 1987A versus time since explosion, with a sigmoid function fit overplotted as the blue dotted line. From Wesson et al. 2015 \citep{Wesson2015}.}
\label{Fig:wesson15}
\end{figure}

The factor of 1000 increase in SN~1987A's ejecta dust mass between day 775 (year 2.1) to years 23-28 after explosion is striking. Gall et al. \citep{Gall2014} combined two published dust mass measurements of SN~1987A with dust masses from infrared SED fits for other CCNSe, obtained by themselves and others, to produce a plot of dust mass versus post-explosion time that showed steadily increasing dust masses over time for these objects. Wesson et al. \citep{Wesson2015} produced a similar dust mass plot for multi-epoch SED observations of SN~1987A alone, which they fitted with a sigmoid curve that reached a dust mass of $\sim$1.0~M$_\odot$ at very late epochs - see Fig.~\ref{Fig:wesson15}. The gap in infrared observations of SN~1987A between day 1153 (year 3.2) and year 23 was partially bridged by Monte Carlo radiative transfer modelling 
\citep{Bevan2016} of the H$\alpha$ and [O~{\sc i}] 6300,6363~\AA\ line profile asymmetries in optical spectra of SN~1987A, which yielded ejecta dust masses at nine epochs between day 714 (year 1.95) and day 3604 (year 9.9)\footnote{Beyond day 3604, optical line emission from the ejecta was obscured by strong emission from the ejecta interaction with the ER}. A sigmoid curve fit to the SN~1987A dust masses obtained only from optical line profile fits up to day 3604 also implied that a dust mass of $\sim$1.0~M$_\odot$ would be reached at very late epochs \citep{Bevan2016}. All three of the above studies \citep{Gall2014, Wesson2015, Bevan2016} inferred large grain radii ($>$0.5~$\mu$m) at late epochs, implying a significant growth in grain size over time. For an approximately constant number of grain nucleation sites, a factor of 1000 increase in overall dust mass between years 2 and 25 would be consistent with a factor of 10 increase in mean grain radius over that time period.

The most recent study of CCSN ejecta dust masses as a function of time since explosion \citep{Niculescu-Duvaz2022} combined a large number of IR SED-based and optical line profile fitting-based dust masses
for CCSNe observed up to sixty years after outburst with those for SN~1987A and for several core-collapse supernova remnants less than a thousand years old.  A sigmoid curve fit to the dust masses obtained for 25 CCSNe and 6 CCSNRs observed over a wide range of epochs yielded a saturation dust mass of 0.42$^{+0.09}_{-0.05}$~M$_\odot$,implying that SN~1987A's dust yield is similar to that of other CCSNe.

Although CCSN dust formation models \citep[e.g.][]{Kozasa1991, Todini2001, 
Sarangi2013} had predicted that up to half a solar mass (170,000 Earth 
masses) of dust could form after a supernova event, they also predicted 
that the timescale for completion of the dust formation process should be 
no more than 3 to 5 years \citep[][]{Sarangi2015, Sluder2018, 
Sarangi2022}, rather than the two decades observed for SN~1987A. Dwek et 
al. \citep{Dwek2015, Dwek2019} have proposed that the apparent discrepancy 
between the predicted and observed dust formation timescales could be 
resolved if almost all of the dust had actually formed during the first 
two years after outburst, as predicted by most CCSN dust condensation 
models, but that the dust was initially very optically thick at IR 
wavelengths, with the observed early IR emission originating from an 
optically thin infrared `photosphere' in the ejecta. Their modelling 
showed that more and more of the dust mass would be revealed as the 
expanding ejecta reduced its optical depth at infrared wavelengths as time 
progressed.

While the above interpretation focused on dust masses derived from analyses of infrared SEDs, Wesson \& Bevan \citep{Wesson2021} investigated how this scenario would affect dust masses determined at similar epochs from both IR SED fitting and from the fitting of optical emission line red-blue profile asymmetries. They confirmed that for early epochs it would be possible to fit the observed IR SEDs with much larger masses of dust located in optically thick clumps, compared to when IR-optically thin dust distributions were assumed. However, they also found that for such large clumped dust masses it was then not possible to fit the optical line profile asymmetries observed at the same epoch. Conversely, clump geometries that could reproduce the observed line profile asymmetries with large dust masses could not reproduce the observed SEDs. In contrast, they showed that at day $\sim$800 both the observed optical-IR SED and the optical line profile asymmetries of SN~1987A could be matched with a low dust mass of $\sim10^{-3}$~M$_\odot$ at that epoch.

\subsubsection{Dust formation in SN~1987A's ejecta - summary}

The SED of the 0.4-0.6~M$_\odot$ of dust estimated to be present in the ejecta of SN~1987A by year 25 after outburst peaked then at $\sim$170~$\mu$m - see Figure~8. None of the  far-infrared space missions currently proposed would have the ability to detect a similar source at that wavelength beyond M~31 in the Local Group, while ALMA's 440-$\mu$m detection of SN~1987A at 50~kpc is near its current detection limit. The peak of SN~1987A's cooling dust SED moved beyond 30~$\mu$m within 3 to 5 years after outburst, so even with {\em JWST's} superb 1-28-$\mu$m sensitivity, observations by it of a similar source to SN~1987A after a similar time period would be unable to measure the total quantity of dust that had actually formed. 

An alternative dust mass measurement technique that still works at late and very late epochs is the measurement and fitting of red-blue asymmetries in optical emission lines. These are caused by photons from the far side of the ejecta experiencing more absorption and scattering by dust as they traverse the ejecta than do blue-shifted photons emitted from the near side of the ejecta. This method is insensitive to the temperature of the dust but does rely on the ejecta being sufficiently illuminated to emit a detectable optical spectrum. This illumination can either be from irradiation of the ejecta by X-ray and UV photons from shocked circumstellar material (CSM), or via excitation by a reverse shock created by the ejecta-CSM collision that is propagating back into the ejecta. Currently the optical to mid-infrared spectrum of SN~1987A is dominated by emission from the shocked ER but once the ER dust, and most of the ER itself, has been destroyed, emission from the ejecta should begin to dominate its spectrum.

As we have seen, due to its proximity, along with its detection and the entirety of its subsequent evolution having taken place during the era of modern technology, SN~1987A has become the archetype for the study of dust formation in CCSN ejecta. During the first 25 years after outburst the dust in its ejecta cooled and grew in mass to between 0.4-0.6~M$_\odot$, a value close to the limit imposed by the total mass of condensable heavy elements predicted to be ejected by a typical current CCSN event. The current SN~1987A dust mass is similar to values that have been derived for other CCSNe from red-blue optical emission line asymmetries measured at similarly late epochs. 

CCSNe appear to be the only plausible source for the dust detected in some very high redshift galaxies, due to the very short evolutionary timescales implied. By analogy to SN~1987A, for the first generation of massive stars a dust yield per CCSN nearly equal to the total mass of condensable heavy elements predicted to be formed and ejected by such objects would appear to be realistic.

\subsection{Molecules in Supernova 1987A's ejecta}

Molecules provide the building blocks for dust grain formation, so determining the relative abundances of different molecular species can help guide us as to what types of dust grains may have formed in the ejecta (e.g. carbon-rich or oxygen-rich). The differing masses of different atomic isotopes within a molecule (e.g. $^{28}$Si, $^{29}$Si or $^{30}$Si in SiO molecules), as well as nuclear hyperfine splitting effects in some cases, lead to easily observable velocity offsets between the same submillimetre rotational transition of different isotopic variants, enabling nucleosynthetic processes that took place in the star, prior to, or during the explosion, to be diagnosed from the relative abundances of the different isotopic variants of a molecule.
Finally, the low excitation energies of molecular rotational transitions make them suitable for determining physical conditions (e.g. temperature, density) in inner ejecta regions, where we know from submillimetre dust observations that the gas and dust may have cooled to temperatures as low as 20~K.

\subsubsection{Early observations of CO and SiO molecules}

Molecules were detected in the ejecta of SN~1987A quite soon after the explosion event. From day 112 onwards, near-IR emission in the 2.3-$\mu$m $v=2-1$ and 4.6-$\mu$m $v=1-0$ vibrational bands of CO was detected \citep{Spyromilio1988, Oliva1987}. Liu et al. \citep{Liu1992} analysed the various observations of CO made between days 192 and 377 and derived a fairly constant CO mass of $10^{-3}$~M$_\odot$ during that period. They found that their chemical modelling could reproduce this mass of CO only if it resided in clumps of microscopically unmixed gas occupying at least 10\% of the volume of the ejecta. The near-IR CO emission had faded below detection limits by day 600, interpreted by \citep{Liu1995} as due to the ejecta gas becoming too cool to excite the vibrational bands of CO.

On day 260 the 8-$\mu$m $v=1-0$ fundamental band of silicon monoxide was detected in emission but had disappeared by day~578\citep{Wooden1993}. The mass of SiO molecules at day 500 was estimated to be 4$\times10^{-6}$~M$_\odot$ \citep{Roche1991}, about 10\% of the dust mass inferred from the day~600 mid-IR continuum emission \citep{Moseley1989, Wooden1993}. Roche et al. \citep{Roche1991} noted that the day~500 SiO excitation temperature of $\sim$1500~K was close to the condensation temperature of silicates, so the non-detection of the 8-$\mu$m SiO band by day 578 may have been due to the incorporation of the majority of SiO molecules into dust grains by that date. The increasing strength of the mid-IR dust continuum by that time would also have made the detection of SiO band emission more difficult.

\subsubsection{Late-time observations of CO, SiO and other molecules in the ejecta}

Kamenetzky et al. \citep{Kamenetzky2013} reported ALMA observations of SN~1987A obtained in 2012 (year 25) that detected the pure rotational J = 2-1 and 1-0 lines of CO, at wavelengths of 2.6 and 1.3~mm, respectively. When combined with {\em Herschel} SPIRE detections of the shorter wavelength J = 7-6 and 6-5 lines of CO, also obtained in 2012, their modelling implied a CO temperature $>10$~K and a total CO mass of $> 0.01$~M$_\odot$, i.e. the CO mass had grown by at least an order of magnitude since day $\sim$400. ALMA's 0.6~arcsec angular resolution used for the CO 2-1 line observation showed that the CO emission was unresolved and confined well inside the ER, consistent with a location deep within the ejecta. The CO 2-1 line's FWHM of 2200~km~s$^{-1}$ was also consistent with an origin in the inner parts of the ejecta.


\begin{figure}
\centering
\includegraphics[width=0.75\hsize]{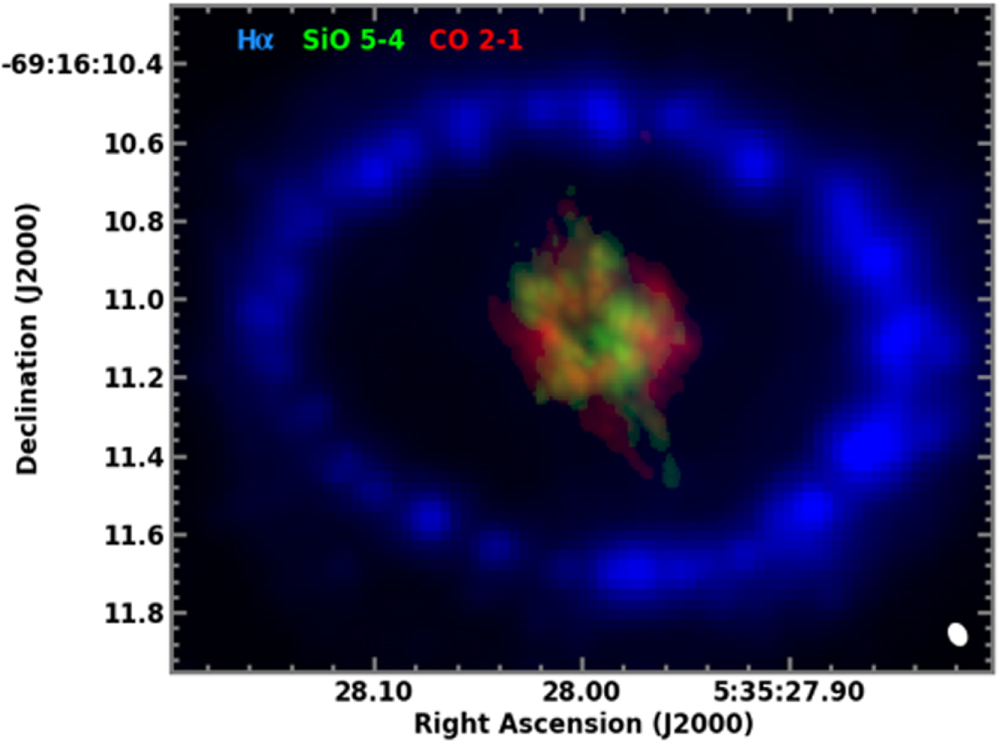}
\caption{The emission at the centre of the image corresponds to the peak intensities of CO 2-1 (red) and SiO 5-4 (green) observed with ALMA. The surrounding H$\alpha$ emission (blue), observed with the {\em HST}, shows the location of the circumstellar equatorial ring. From Abellan et al. 2017 \citep{Abellan2017}.}
\label{Fig:abellan17_image}
\end{figure}

In deeper ALMA 210-360~GHz spectra, obtained in 2014 
\citep{Matsuura2017},
the SiO J=5-4 and 6-5 lines, the 2-1 and 3-2 lines of CO, the J=3-2 line of HCO$^+$ and the J$_K$ = 6$_6$-5$_5$ line of SO  were all detected, with FWHM line widths between 1770 and 2210~km~s$^{-1}$. Combining their CO line data with those of Kamenetzky et al. discussed above, their modelling suggested particle densities in the CO-emitting region of between 10$^5$ and 10$^6$~cm$^{-3}$ and gas kinetic temperatures of between 20 and 50~K, with a minimum CO mass of 0.02~M$_\odot$ \citep{Matsuura2017}. Their results confirmed that, as predicted \citep{Liu1992}, strong cooling of the inner ejecta gas had taken place since the near-IR vibrational lines of CO had last been detected, several hundred days after outburst. 

\begin{figure}
\centering
\includegraphics[width=0.6\hsize]{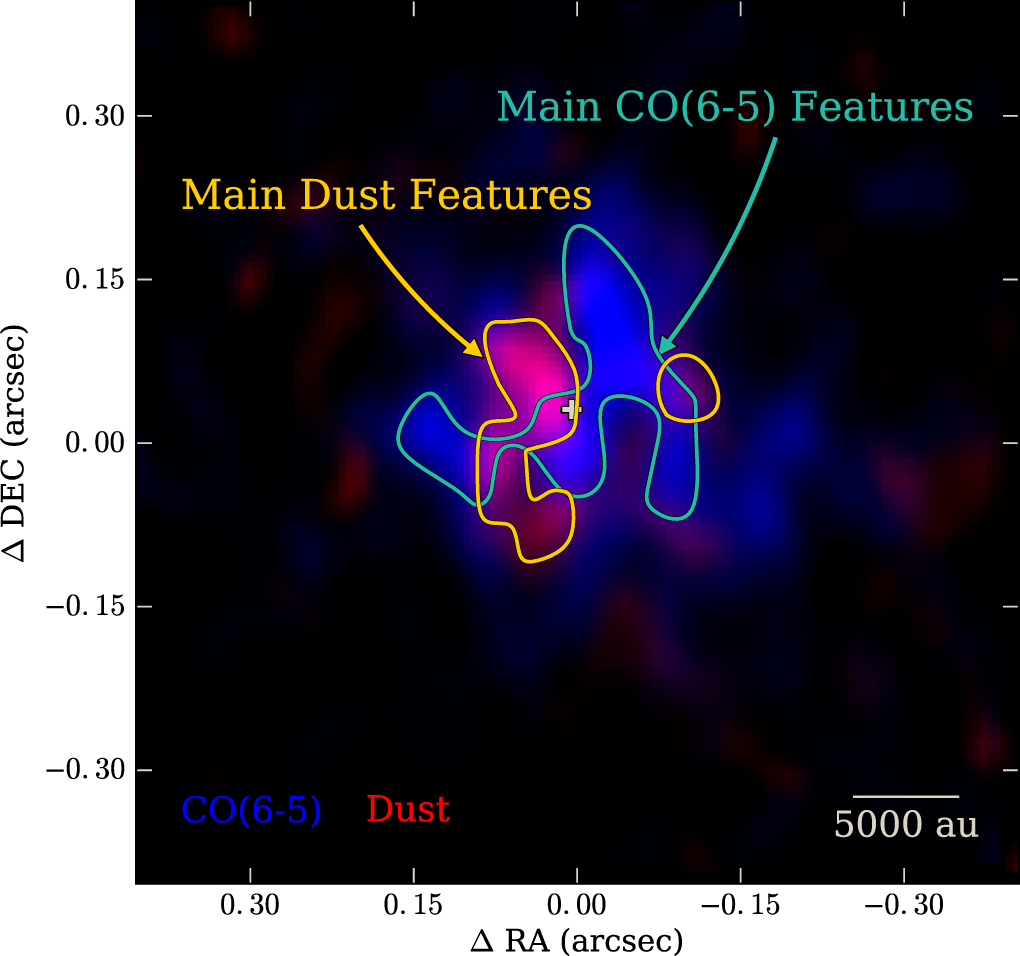}
\caption{Illustration of the spatial anti-correlation between the dust emission peak at 442~$\mu$m (red) and the CO J = 6-5 emission (blue), both measured by ALMA. The plus sign denotes the system centre position.
From Cigan et al. 2019 \citep{Cigan2019}.}
\label{Fig:Cigan19_fig4}
\end{figure}

The above deep 2014 and 2015 ALMA observations of the CO 2-1 and SiO 5-4 rotational lines were acquired with an angular resolution of 50 milliarcseconds. The spatial distribution of the emission from both lines (see Fig.~\ref{Fig:abellan17_image}) was found to be largely confined within a region of angular diameter 0.6~arcsec, to have a toroidal or shell-like structure and to be highly clumped, with the CO emission extending further out than the SiO emission. These properties were interpreted as confirming that non-spherical instabilities had been present at the time of the explosion \citep{Abellan2017}.

Cigan et al. \citep{Cigan2019} presented ALMA line and continuum data on SN~1987A acquired in the 1300-, 870- and 450-$\mu$m spectral windows (Bands 6, 7 and 9) during the second half of 2015 (year 28.5). Their coverage included the J=2-1, 3-2 and 6-5 rotational lines of CO and the J=5-4, 6-5, 7-6 and 8-7 lines of SiO (although the SiO 8-7 line was heavily blended with the CO J=3-2 line so that neither of those two lines could be used). The ALMA data also enabled dust continuum and emission line images to be generated across these bands. Fig.~\ref{Fig:Cigan19_fig4} \citep{Cigan2019} compares the distribution of the CO 6-5 line emission with that of the dust continuum emission at a nearby wavelength and shows that both the CO emission and the dust emission are clumpy, with the brightest dust emission located in regions of relatively faint CO emission and vice versa.

\subsubsection{Molecular hydrogen in the ejecta}

Molecular hydrogen emission was not detected from SN~1987A during the first several hundred days after outburst, despite having been predicted to form in the ejecta at early epochs, along with CO and SiO molecules \citep{Culhane1995}.
However, 2.12-$\mu$m and 2.40-$\mu$m $v=1-0$ vibration-rotation line emission from molecular hydrogen located in the ejecta was eventually detected from day 6489 onwards \citep{Fransson2016}, the first time that H$_2$ had been detected in the ejecta of any supernova.
The radioactive decay of $^{44}$Ti, rather than external irradiation by UV or X-ray photons from the ER interaction regions, was favoured as the energy source for the H$_2$ line excitation, due to the relative lack of variation in the H$_2$ line fluxes over a nine-year period.  

The H$_2$ emission was found to be dominated by a single clump in the southern ejecta, which on days 8694-11275 had a central velocity of $\sim$1700~km~s$^{-1}$ and a FWHM of $\sim$1400~km~s$^{-1}$ \citep{Larsson2019a}. Observed H$_2$ velocities across the surface of the ejecta ranged as low as 400-800~km~s$^{-1}$, slightly lower than seen for CO or SiO, consistent with significant inward mixing of hydrogen and providing direct evidence for mixing between the different nuclear burning zones during the explosion \citep{Larsson2019a, McCray1993}.

The {\em JWST} NIRSpec 1-5-$\mu$m IFU spectrum of the inner ejecta region \citep{Larsson2023} detected many vibration-rotation emission lines of H$_2$. A good match to the observed H$_2$ spectrum was obtained using a photodissociation region (PDR) model having a high density and a high far-UV flux, with UV fluorescence as the main source of excitation for the H$_2$. By the 2022 epoch of the NIRSpec observations, the increasing irradiation of the ejecta by X-rays and UV photons from the ER interaction regions may have supplanted $^{44}$Ti decay as the main power source for the H$_2$ line emission.

\subsection{Simulations of the SN~1987A explosion and the formation of asymmetric ejecta}

The X-ray and $\gamma$-ray emission from SN~1987A has been modelled using a spectral synthesis code that was applied to 3D core collapse models evolved to late times
\citep{Alp2019, Jerkstrand2020}, with particular emphasis in the latter work on the emergent line profiles of $\gamma$-ray decay lines from $^{56}$Co and $^{44}$Ti. For neutrino-driven explosions of evolving single stars having initial main sequence masses of 15-20~M$_\odot$, explosion energies of $\sim1.5\times10^{51}$~ergs produced ejecta structures that could match the observed $\gamma$-ray line velocity profiles and which implied a neutron star kick of at least 500~km~s$^{-1}$.
They also found that in order to reproduce the UV-Optical-IR bolometric light curve up to day 600, a total ejecta mass of $\sim$14~M$_\odot$ was needed \citep{Jerkstrand2020}.

Orlando et al. \citep{Orlando2020} also found that neutrino-driven CCSN explosions can in some cases result in strongly bipolar ejecta morphologies \citep{Wongwathanarat2015, Ono2020, Wang2024, Vartanyan2025}. Starting with a pre-SN stellar model of an 18.3~M$_\odot$ blue supergiant (BSG) produced by the merger of 14 and 9~M$_\odot$ stars \citep{Urushibata2018}\footnote{For their own 3D hydrodynamical simulations, Ono et al. \citep{Ono2020} also found this BSG starting model to provide the best fit to observations.} they found that this ultimately provided a better match to observations than single star models. Using this BSG model, they carried out a simulation (based on \citep{Ono2020}) of the initial 20 hours after core collapse, This in turn served as the starting point for 3D MHD simulations that treated the subsequent expansion of the remnant for the next 50 years through a pre-SN environment (1) with an $r^{-2}$ radial density distribution produced by the supergiant progenitor, and (2) with structures similar to the triple-ring system around SN~1987A. Their best-fitting models to early infrared emission line profiles required an explosion energy of $2\times10^{51}$~ergs and a high degree of asymmetry, with the radial velocity along the polar axis
16 times greater than the radial velocity in the equatorial plane and with the polar axis of the ejecta lying nearly in the plane of the ER. They also used the distributions of $^{12}$C, $^{16}$O and $^{28}$Si in their stellar models to map the distributions of the square roots of the products CxO and SixO, as proxies for the CO and SiO molecular emission distributions mapped by ALMA in the core of the ejecta. CxO was found to be more extended than SixO in the models \citep{Orlando2020, Ono2024}, in agreement with the ALMA observations of CO and SiO \citep{Abellan2017}. 

\begin{figure}
\centering
\includegraphics[width=0.7\hsize]{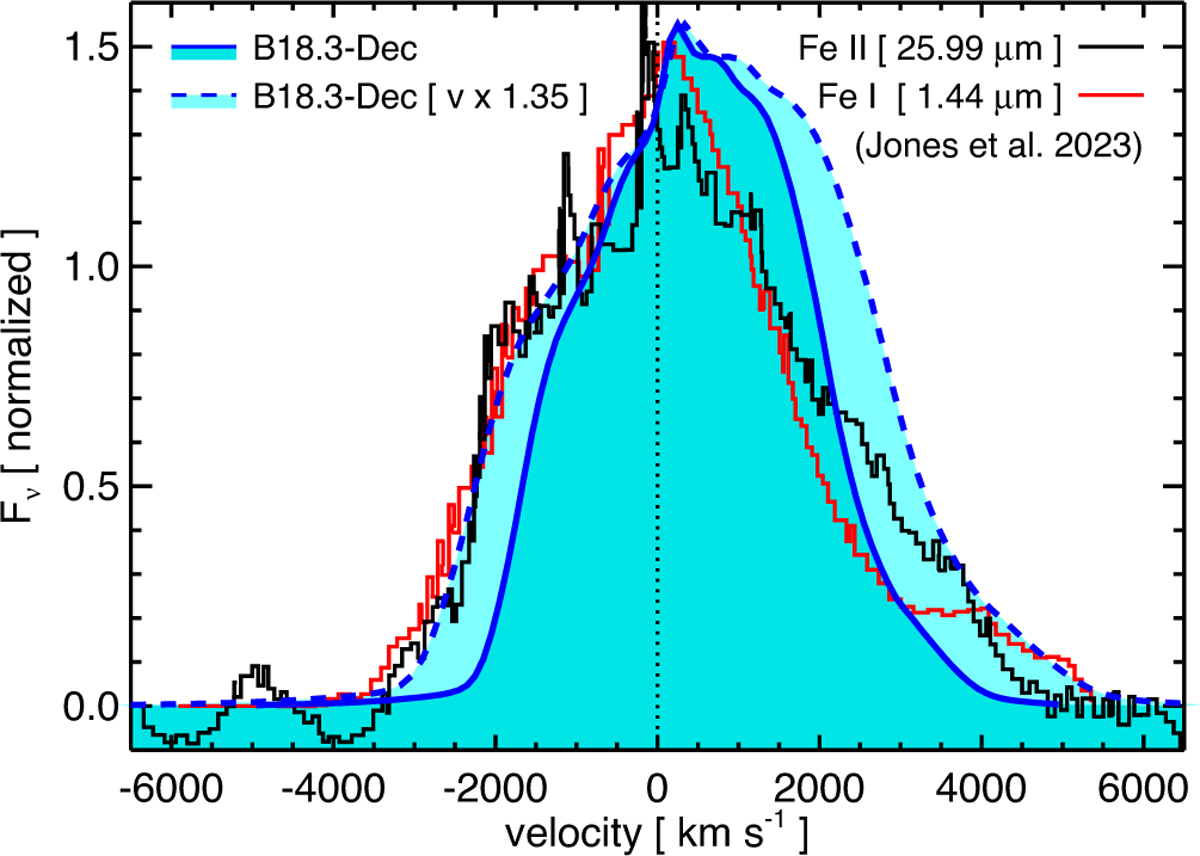}
\caption{Profiles of continuum-subtracted [Fe~{\sc ii}] 25.99~$\mu$m (black line, from \citep{Jones2023b})
and [Fe~{\sc i}] 1.443~$\mu$m (red line, from \citep{Larsson2023}) emission lines from the inner ejecta of SN~1987A, observed on July 16, 2022 with {\em JWST}. These are compared with the total mass distribution of unshocked Fe-rich ejecta as a function of the line of sight (LoS) velocity predicted for February 2023 by model B18.3-Dec of \citep{Orlando2025} (solid blue line). The dashed blue line represents the Fe mass distribution with the LoS velocity artificially increased by 35\% to better match the {\em JWST} observations. The Fe mass distributions were normalized to roughly match the peak of the [Fe~{\sc i}] line.  From Orlando et al. 2025 \citep{Orlando2025}.
}
\label{Fig:orlando25}
\end{figure}

Subsequent hydrodynamical simulations \citep{Orlando2025}, also using as a starting point the 18.3~M$_\odot$ binary-merger blue supergiant stellar model, 
reproduced the  morphologies and velocities of [Fe~{\sc i}] and [Fe~{\sc ii}] emission lines seen in the {\em JWST} NIRSpec and MIRI-MRS spectra of the ejecta (see  Fig.~\ref{Fig:orlando25}) although some remaining discrepancies in clump velocities and spatial distributions suggested even stronger asymmetries during the explosion than had been modelled.

1.7-10~keV X-ray spectra of SN~1987A, obtained in June 2024 by the {\em XRISM}
observatory, revealed heavy element emission lines with line widths of 1500-1700~km~s$^{-1}$ \citep{Audard2025}, similar to the expansion velocities found at other wavelengths for regions in the main ejecta. Their model analysis indicated that the X-ray lines originate from `non-metal-rich' regions with abundance patterns characteristic of the LMC, interpreted as originating from material that was originally in the outer envelope of the progenitor star and which is now at a temperature of 3.3$\times$10$^7$~K after  encountering the reverse shock 
generated by the impact of the ejecta on the ER \citep{Audard2025}, a result in good agreement with published predictions \citep{Orlando2020, Sapienza2024}.

\section{The long search for the remnant neutron star}

\subsection{Background}

The detection of a burst of neutrinos from SN~1987A provided overwhelming evidence that a neutron star had formed as a result of the core collapse, but the possibility always existed that infalling material could have triggered a subsequent collapse to a black hole. Over the years a variety of searches for evidence for the continued existence of a neutron star have been made, as most recently reviewed in 2007 \citep{McCray2007} and in 2018 \citep{Alp2018}. These searches ranged across the electromagnetic spectrum, from gamma-ray to radio wavelengths, but yielded only upper limits to the luminosity of a neutron star or pulsar wind nebula \citep{Alp2018, Alp2021, Dohi2023}. Radio pulsar searches have also only been able to set upper limits (e.g. \citep{Zhang2018}).

From their 2015 ALMA observations of SN~1987A at 679~GHz (442~$\mu$m), Cigan et al. \citep{Cigan2019} found a bright dust emission peak, which they called `the blob' (see Fig.~\ref{Fig:Cigan19_fig4}) located 84~milliarcseconds to the northeast of the position of the progenitor star that had been measured by Alp et al. \citep{Alp2018}. This brightness peak was attributed to dust heated by a compact object, e.g. a pulsar wind nebula or cooling neutron star \citep{Cigan2019}, with a cooling neutron star favoured by \citep{Page2020}.
An 84~milliarcsecond displacement between 1987 and 2015 from the pre-explosion position of the progenitor star  would correspond to a space velocity of 700~km~s$^{-1}$, (just) consistent with the range of neutron star kick velocities estimated for a number of other supernova remnants known to host compact objects \cite{Zanardo2014}.

In 2021 Greco et al. \citep{Greco2021} published a re-analysis of archival {\em Chandra} and {\em NuSTAR} 2012-2014 X-ray observations of SN~1987A. In addition to a soft X-ray component associated with shocked circumstellar material, a second, hard X-ray, component was evident in the {\em NuSTAR} data at energies above 10~keV. This was attributed to nonthermal synchrotron emission, either from diffusive shock acceleration in the outer layers of the ejecta, or from a pulsar wind nebula (PWN) embedded at the centre of the ejecta, with parameters consistent with other members of the PWN population.

\subsection{The detection of a central compact object using the {\em JWST}}

\begin{figure}
\centering
\includegraphics[width=\hsize]{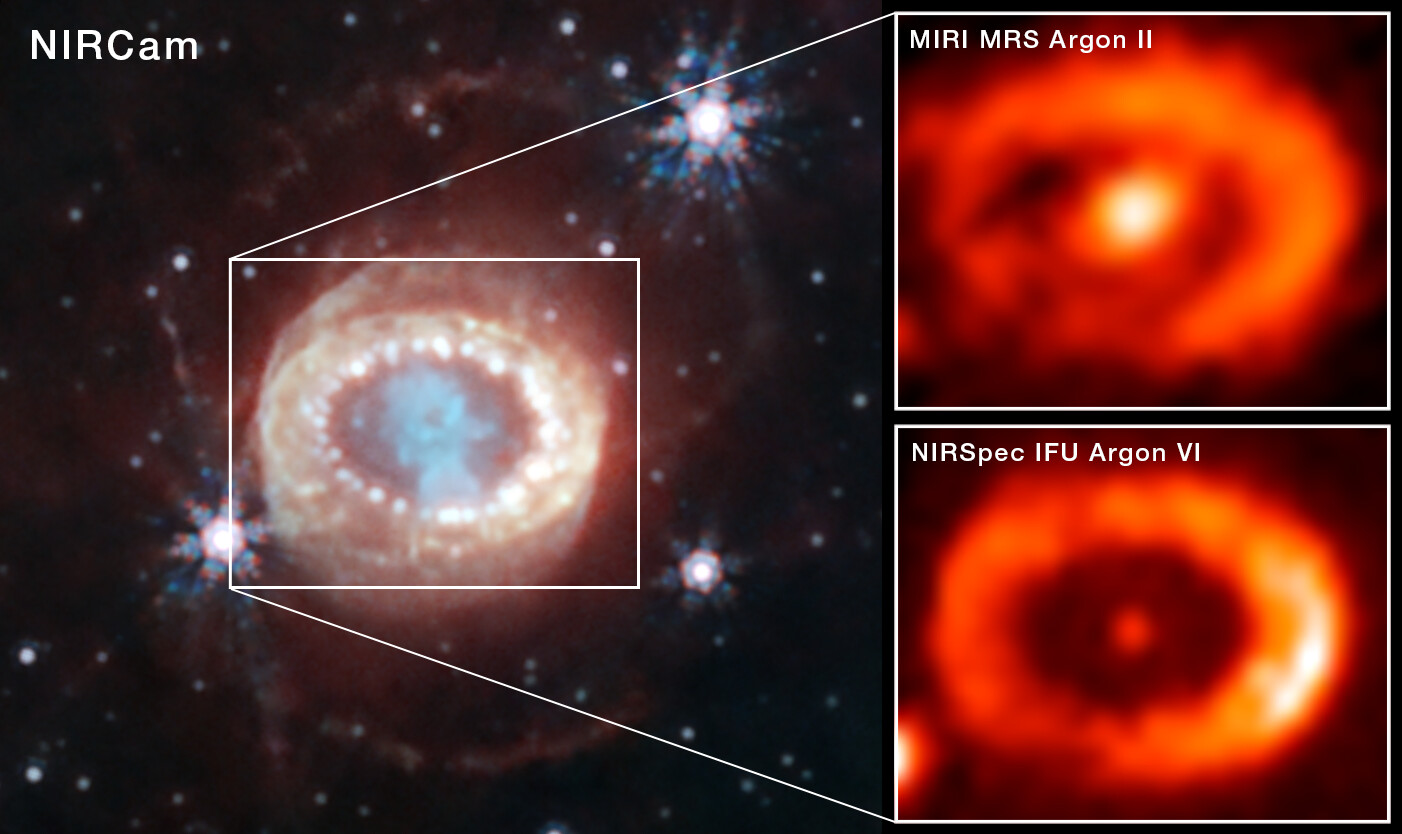}
\caption{Left: The {\em JWST} NIRCam near-infrared composite image of the SN~1987A remnant shown in Fig.~\ref{Fig:87A_nircam}; Right (top): MIRI-MRS IFU image at the wavelength of the [Ar~{\sc ii}] 6.985~$\mu$m transition, showing strong line emission at the centre of the inner ejecta; Right (bottom): NIRSpec IFU image at the wavelength of the [Ar~{\sc vi}] 4.529~$\mu$m transition, showing weaker line emission from the ejecta centre. In both the MIRI-MRS and NIRSpec images \citep{Fransson2024} the emission from the equatorial ring is dominated by continuum processes.
Image credit: NASA, ESA, CSA, C. Fransson, M. Matsuura et al.}
\label{Fig:esa_publ}
\end{figure}

\begin{figure}
\centering
\includegraphics[width=0.75\hsize]{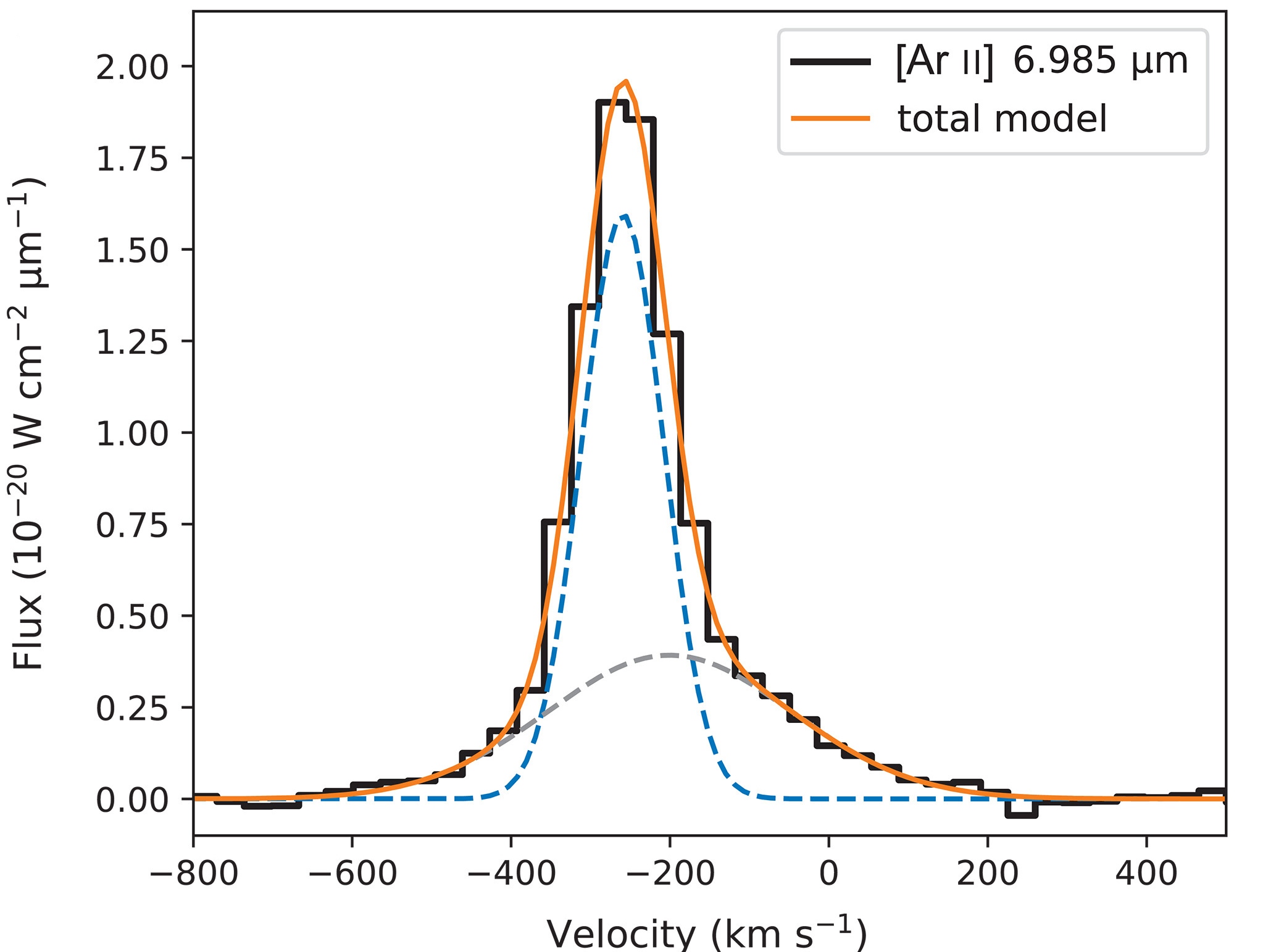}
\caption{{\em JWST} MIRI-MRS velocity profile of the [Ar~{\sc ii}] 6.985-$\mu$m line from the central ejecta of SN~1987A (black histogram). The orange curve is a model fitted to the data, consisting of two Gaussian components (blue and grey dashed curves), which are both blue-shifted, by -259.6~km~s$^{-1}$ and -200.6~km~s$^{-1}$, respectively, relative to the systemic radial velocity of SN~1987A at zero km~s$^{-1}$. From Fransson et al. 2024 \citep{Fransson2024}.
}
\label{Fig:ArII_profile}
\end{figure}

The {\em JWST} Cycle~1 observations of SN~1987A, obtained on day 12,927 
with the 5-28-$\mu$m MIRI-MRS IFU, unexpectedly revealed strong [Ar~{\sc ii}] 6.985-$\mu$m forbidden line emission coming from the very centre of the ejecta \citep{Fransson2024}, see Fig.~\ref{Fig:esa_publ}. Inspection of 1-5-$\mu$m NIRSpec IFU observations of SN~1987A obtained on the same date revealed weaker emission in 
the [Ar~{\sc vi}] 4.529-$\mu$m forbidden line at the same location. The velocity profile of the [Ar~{\sc ii}] line is displayed in Fig.~\ref{Fig:ArII_profile}. It shows two components: one displaced by -259.6$\pm$0.4~km~s$^{-1}$ from the rest frame of SN~1987A\footnote{The rest frame of SN~1987A corresponds to a heliocentric velocity of +286.5~km~s$^{-1}$ \citep{Groningsson2008}}, with a FWHM of 121.9$\pm$1.3~km~s$^{-1}$, close to the instrumental resolution of the MRS at that wavelength; and the other displaced by -200.6$\pm$3.7~km~s$^{-1}$ from the SN~1987A rest frame, with a FWHM of 362.4$\pm$8.0~km~s$^{-1}$. 

The Cycle~1 NIRSpec IFU observation of the [Ar~{\sc vi}] 4.529-$\mu$m line was obtained with a factor of three lower spectral resolving power than the [Ar~{\sc ii}] MRS observations but could also be fitted by two components, with the narrower component displaced by -269$\pm$24~km~s$^{-1}$ from the SN~1987A rest frame \citep{Fransson2024}. The [Ar~{\sc ii}] and [Ar~{\sc vi}] central emission peaks were both spatially unresolved but coincided spatially within the measurement errors, while the higher diffraction-limited spatial resolution of the NIRSpec IFU at 4.529~$\mu$m (see Fig.~\ref{Fig:esa_publ}) allowed the centroid of the [Ar~{\sc vi}] emission peak to be determined to be 38$\pm$22~milliarcseconds (mas) east and 31$\pm$22~mas south of the geometric centre of the ER that had been previously determined from {\em HST} optical images \citep{Alp2018}. These spatial offsets of the [Ar~{\sc vi}] emission peak, combined with the -259.6$\pm$0.4~km~s$^{-1}$ radial velocity of the main blue-shifted [Ar~{\sc ii}] component, correspond to a 3D space velocity of 416$\pm$206~km~s$^{-1}$ since the explosion event \citep{Fransson2024}. Such kicks are usually attributed to asymmetries in the original explosion (e.g. \citep{Janka2024}).

The position of the brightest pixel of the ALMA dust emission peak, at 72~mas east and 44~mas north of the geometric centre of the ER \citep{Cigan2019}, was noticeably offset from the [Ar~{\sc vi]} emission peak 38$\pm$22~mas east and 31$\pm$22~mas south
of the centre.

In addition to the detection of emission from two ionized species of argon, the 
Cycle~1 MRS observations also detected narrow [S~{\sc iv}] 10.51-$\mu$m and [S~{\sc 
iii}] 18.71-$\mu$m line emission from the centre of the ejecta, with blue-shifted 
radial velocities similar to those of the argon lines. Repeat MRS observations in 
Cycle~2 led to the detection of [Cl~{\sc ii}] 14.468-$\mu$m emission from the ejecta 
centre, with a blue-shifted radial velocity of -304$\pm$14~km~s$^{-1}$ 
\citep{Larsson2025}, while higher spectral resolution NIRSpec IFU observations in 
Cycle~2 confirmed the presence of two central [Ar~{\sc vi}] 4.529-$\mu$m velocity 
components, with radial velocities and line widths similar to those of the [Ar~{\sc 
ii}] 6.985-$\mu$m line. In addition, the higher resolution NIRspec spectra 
enabled the 
detection of similarly blue-shifted velocity components in the 4.159-$\mu$m line of 
[Ca~{\sc v}] and the 1.644-$\mu$m line of [Fe~{\sc ii}] \citep{Larsson2025}.

All of the elements whose lines show narrow blue-shifted velocity components at the 
centre of the ejecta (S, Cl, Ar, Ca, Fe) originate from oxygen- and silicon-burning 
inner nucleosynthetic layers of the star. whereas lighter elements with accessible 
ionic fine-structure lines, such as oxygen and neon, which should occur further out 
in the ejecta, do not exhibit these narrow blue-shifted velocity components.

In 1992 Chevalier \& Fransson \citep{Chevalier1992} had predicted that a pulsar nebula within the SN~1987A remnant could generate a high enough ionizing photon luminosity to excite emission lines from heavy elements deep in the ejecta. In order to account for the {\em JWST} observations, Fransson et al. \citep{Fransson2024} constructed models in which the inner ejecta is being photoionized by two alternative radiation sources: (a) a hot cooling neutron star (CNS) with a temperature $(1.5-3)\times10^6$~K and an ionizing photon luminosity $L_{\rm ion} = 78~L_\odot$; or (b) a pulsar wind nebula, produced by a neutron star at the centre, having a Crab Nebula-like spectrum with $L_{\rm ion} = 13~L_\odot$. Both models had similar ionization parameters, $L_{\rm ion}/n_{\rm ion}r^2$, where $r$ is the distance to the ionizing source and $n_{\rm ion}$ is the number density of ions at the inner edge of the ejecta, with $n_{\rm ion} = 2.6\times10^4$~cm$^{-3}$ and a volume filling factor of 0.1 providing the best fit to line ratios. In order to obtain a match to the fluxes or flux upper limits of some lines, particularly those in the 8-13-$\mu$m and 18-$\mu$m regions, it was necessary to introduce an overlying silicate dust extinction curve corresponding to an optical depth $\tau_{\rm abs}= 6.5$ at a wavelength of 10~$\mu$m \citep{Fransson2024}. 
We know from the ALMA submillimetre observations of SN~1987A discussed in Section~3.1.2,
that large quantities of dust are present in the ejecta, including close to the very centre. Chugai \& Utrobin \citep{Chugai2024} have interpreted the broad [Ar~{\sc ii}] component (FWHM $\sim360$~km~s$^{-1}$) that lies 59~km~s$^{-1}$ to the red of the narrow component (FWHM $\sim120$~km~s$^{-1}$), see Fig.~14, as being caused by the scattering of narrow-line photons off an optically thin inter-clump dust component, requiring grain radii of 1-2~$\mu$m and a dust component mass of a few$\times10^{-3}$~M$_\odot$.

Rosu et al. \citep{Rosu2024} have presented {\em HST}-WFC3 imaging of SN~1987A obtained in 2022 (day 12,980) with the F502N narrow-band filter (FWHM = 65~\AA ), which isolates the [O~{\sc iii}] 5008~\AA\ emission line. For a position at the centre of the ejecta, corresponding to the location of the central [Ar~{\sc ii}] line, they could obtain only an upper limit for the flux in the [O~{\sc iii}] line which was lower than the predictions for both the PWN and CNS models of \citep{Fransson2024}.
For the most likely case, where photon scattering by dust at 5008~\AA\ is very strong and the [O~{\sc iii}] emission is as a result spread over the entire volume of the ejecta, their measured integrated F502N filter flux of $3\times10^{-16}$~ergs~cm$^{-2}$~s$^{-1}$ is 2.5 times smaller than the PWN model prediction but 11 times larger than the CNS model prediction, leaving the PWN model as the less favoured of the two models. 
A caveat to this conclusion however is that the non-detection of the [O~{\sc iii}] line from the centre of the ejecta could indicate that the region from which the observed Ar and S emission lines originate is confined to the innermost S- and Ar-rich zones, where the O abundance is expected to be much lower \citep{Rosu2024}. Spatially resolved imaging spectroscopy
of the ejecta at the wavelengths of the [O~{\sc iii}] doublet could help resolve this issue.

To summarise, observations obtained with the {\em JWST} have provided a crucial breakthrough in the hunt for a remnant compact object in SN~1987A, via the detection, at the very centre of the ejecta, of redshifted mid-infrared line emission from argon and other ions. This emission can only have been powered by the photoionization of material located at the inner edges of the ejecta, by either a hot cooling neutron star or a pulsar wind nebula.

\section{Future prospects}

Over the next two decades the  Supernova 1987A system will evolve further while, in addition to the current complement of spaceborne and ground-based facilities, several 
new and powerful facilities will become available.

\begin{itemize}

\item 
3D MHD models already exist that track the interaction of the ejecta with the surrounding circumstellar material up to 2037 \citep{Orlando2025}. During this period the main body of the expanding ejecta (total mass 14~M$_\odot$ \citep{Jerkstrand2020}) will continue to over-run the equatorial ring and exit beyond it.
Radiation from the shocked ejecta, which is already making a significant contribution to the overall X-ray luminosity of the system, should dominate by 2027. Future X-ray spectra from {\em XRISM} should eventually reveal heavy element abundance ratios more characteristic of enriched inner ejecta regions interacting with the reverse shock, while a time series of such observations can track the expansion and evolution of the ejecta, providing insights into explosion asymmetries \citep{Sapienza2024}. Infrared observations with the {\em JWST}, particularly of [Fe~{\sc ii}] and [Fe~{\sc i}] emission lines, will also probe the evolution of the ejecta and its interaction with the ring system.

\item 
For radio wavelength observations, the Square Kilometre Array (SKA) is expected to become available from about 2030. The SKA-Mid array, located in South Africa, will have 133 15-m dishes and 64 13.5-m dishes, operating at frequencies between 350~MHz (90~cm) and 15.4~GHz (2~cm). Half the dishes will be concentrated within 2~km, with the others distributed in three spiral arms, with baselines of up to 150~km
and angular resolutions from 0.03 to 1.4 arcseconds.
The SKA-Low array, located in Western Australia, will have 131,000 dipole antennas, operating at frequencies between 350~MHz and 50~MHz (600~cm)
and distributed across 512 stations having 250 antennas each.
Half of the stations will be located within a 1~km diameter core, with the remaining stations organised in clusters of 6 stations distributed on three spiral arms, with baselines of up to 75~km and angular resolutions from 3.3 to 23 arcseconds. Observations of SN~1987A with SKA-Mid will provide a sensitive high angular resolution probe of the non-thermal emission processes at work in the forward and reverse shocks in the system. The huge frequency range covered by the SKA-Low and SKA-Mid facilities, coupled with their unprecedented sensitivities, will enable very deep searches to be made for a radio pulsar in SN~1987A.

\item 
The evidence for the presence of a cooling hot neutron star or pulsar wind nebula in SN~1987A was provided by the detection by the {\em JWST} MRS and NIRSpec spectrometers of narrow redshifted infrared emission lines emerging from the centre of the ejecta. As the ejecta expands, the optical depth of obscuring dust should gradually reduce but the timescale for this is unknown.
Future {\em JWST} observations can monitor emergent line fluxes and profiles, along with the possible appearance of additional lines.
ESO's 39.3-m Extremely Large Telescope (ELT) is expected to have `scientific first light' at the end of 2030, while the 25.4-m Giant Magellan Telescope (GMT), also located in Chile, is expected to commence science operations in the early- to mid-2030's. The adaptive optics of these telescopes will enable optical angular resolutions of 5~milliarcseconds (mas) to be achieved. The Harmoni IFU spectrometer on the ELT will cover the 0.47-2.45-$\mu$m wavelength range with spectral resolving powers of up to 18,000. The ELT's METIS instrument will cover the 3-5-$\mu$m and 7.5-13.5-$\mu$m ranges with diffraction-limited angular resolutions of 2~mas at 3~$\mu$m and $\sim$60~mas at 10~$\mu$m. Low-resolution (R=400-1900) long-slit spectrometers will be available in each wavelength band, together with an R=100,000 IFU spectrometer for the 3-5-$\mu$m band.
The combination of high angular resolution, the high spectral resolution out to a wavelength of 5~$\mu$m, and the huge telescope collecting area, will make Harmoni and METIS powerful probes of emission lines emerging from the central regions of SN~1987A's ejecta, and of shock emission processes in the equatorial ring.

\item
For the photoionization modelling of the central ejecta regions by \citep{Fransson2024},
an ejecta composition typical of explosive oxygen burning in a supernova from a 19~M$_\odot$ progenitor star was adopted, for which the most recently available massive star nucleosynthetic evolutionary calculations are now 18 years old. 
New massive star models, with the most up to date physics, are needed, including models for below-solar initial metallicities. Given the importance of SN~1987A, and the unknown fraction of core-collapse supernovae which result from stars which are the products of stellar mergers, similar models for binary merger systems would be valuable. In particular, a model is needed that computes the evolution of the internal nucleosynthetic structure of a star after it forms from a binary merger event and then explodes after 20,000 years.

\end{itemize}

\bibliographystyle{tfnlm}
\bibliography{87a_rev}{}

\end{document}